\documentclass[11pt,twoside,a4paper,cmspaper,final,collab]{cms-tdr}
\def\svnVersion{6330e30}\def\svnDate{2026/10/03}\def\cmsCernNoTag{CERN-EP-2026-259}\def\cmsCernDate{\today}\def\cmsMessage{Submitted to the Journal of High Energy Physics}
\begin{document}\cmsNoteHeader{EXO-25-001}

\providecommand{\cmsTable}[1]{\resizebox{\textwidth}{!}{#1}}

\newcommand{\Wprime}{\ensuremath{\PWpr}}%
\newcommand{\Zprime}{\ensuremath{\PZpr}}%
\newcommand{\Vprime}{\HepParticle{\PV}{}{\prime}\xspace}%
\newcommand{\tSM}{\ensuremath{{\tan\theta_W}}}%
\newcommand{\cSM}{\ensuremath{{\cos\theta_W}}}%
\newcommand{\sSM}{\ensuremath{{\sin\theta_W}}}%
\newcommand{\Wo}{\PW}%
\newcommand{\Wp}{\PWp}%
\newcommand{\Wm}{\PWm}%
\newcommand{\WW}{\PW\PW}%
\newcommand{\Zo}{\ensuremath{\PZ}}%
\newcommand{\ZH}{\ensuremath{\PZ\PH}}%
\newcommand{\Vo}{\ensuremath{\PV}}%
\newcommand{\MZ}{\ensuremath{m_\PZ}}%
\newcommand{\MW}{\ensuremath{m_\PW}}%
\newcommand{\qqprimebar}{\PQq\PAQq^\prime}%
\newcommand{\eebar}{\Pe^+\Pe^-}%
\newcommand{\mmbar}{\ensuremath{\PGmp\PGmm}}%
\newcommand{\tautaubar}{\ensuremath{\PGtp\PGtm}}%
\newcommand{\llbar}{\ensuremath{\Pellp\Pellm}}%
\newcommand{\lnu}{\ensuremath{\Pell\PGn}}%
\newcommand{\qbar}{\PAQq}%
\newcommand{\ppVll}{\Pp\Pp\to{\Vo}^{(*)}\to\llbar\xspace}%
\newcommand{\ppll}{\Pp\Pp\to\llbar\xspace}%
\newcommand{\ppWln}{\Pp\Pp\to{\Wo}^{(*)}\to\lnu\xspace}%
\newcommand{\ppVjet}{\Pp\Pp\to\Vo^{(*)}+\text{1~jet}\xspace}%
\newcommand{\Vll}{\Vo^{(*)}\to\llbar\xspace}%
\newcommand{\qqVg}{\qqbar\to\Vo\Pg\xspace}%
\newcommand{\qqVgvirt}{\qqbar\to\Vo(\Pg)\xspace}%
\newcommand{\qgVq}{\PQq\Pg\to\Vo\PQq\xspace}%
\newcommand{\qbargVqbar}{\PAQq\Pg\to\Vo\PAQq\xspace}%
\newcommand{\Vb}{\Vo\Pqb\xspace}%
\newcommand{\Vc}{\Vo\Pqc\xspace}%
\newcommand{\chiq}{\chi_{\PQq}\xspace}%
\newcommand{\chil}{\chi_{\Pell}\xspace}%
\newcommand{\chif}{\chi_{\Pf}\xspace}%
\newcommand{\rhon}{\rho_{\mathrm{n}}\xspace}%
\newcommand{\rhoc}{\rho_{\mathrm{c}}\xspace}%
\newcommand{\awn}{a^\mathrm{n}_W\xspace}%
\newcommand{\ayn}{a^\mathrm{n}_Y\xspace}%
\newcommand{\tiq}{T^I_{\PQq}\xspace}%
\newcommand{\til}{T^I_{\Pell}\xspace}%
\newcommand{\ttq}{T^3_{\PQq}\xspace}%
\newcommand{\ttu}{T^3_{\PQu}\xspace}%
\newcommand{\ttd}{T^3_{\PQd}\xspace}%
\newcommand{\ttl}{T^3_{\Pell}\xspace}%
\newcommand{\ttf}{T^3_{\Pf}\xspace}%
\newcommand{\tf}{T^I_{\Pf}\xspace}%
\newcommand{\yq}{Y_{\Pq}\xspace}%
\newcommand{\yl}{Y_{\Pell}\xspace}%
\newcommand{\yf}{Y_{\Pf}\xspace}%
\newcommand{\yyq}{Y_{\chiq}\xspace}%
\newcommand{\yyl}{Y_{\chil}\xspace}%
\newcommand{\yyf}{Y_{\chif}\xspace}%
\newcommand{\Qq}{Q_\PQq\xspace}%
\newcommand{\Ql}{Q_\Pell\xspace}%

\newcommand{\Spar}{\ensuremath{\mathcal{S}}\xspace}%
\newcommand{\Tpar}{\ensuremath{\mathcal{T}}\xspace}%
\newcommand{\Wpar}{\ensuremath{\mathcal{W}}\xspace}%
\newcommand{\Ypar}{\ensuremath{\mathcal{Y}}\xspace}%

\newcommand{\Ho}{\PH}%
\newcommand{\qprimebar}{\ensuremath{\PAQqpr}}%
\newcommand{\gWprime}{\ensuremath{g_{\PWpr}}}%
\newcommand{\gZprime}{\ensuremath{g_{\PZpr}}}%
\newcommand{\gstar}{\ensuremath{g^*}}%
\newcommand{\mstar}{\ensuremath{m^*}}%
\newcommand{\gstarr}{\ensuremath{g^{*2}}}%
\newcommand{\mstarr}{\ensuremath{m^{*2}}}%
\newcommand{\OH}{\ensuremath{\mathcal{O}_{H}}}%
\newcommand{\OW}{\ensuremath{\mathcal{O}_{W}}}%
\newcommand{\OB}{\ensuremath{\mathcal{O}_{B}}}%
\newcommand{\OHW}{\ensuremath{\mathcal{O}_{HW}}}%
\newcommand{\OHB}{\ensuremath{\mathcal{O}_{HB}}}%
\newcommand{\OTW}{\ensuremath{\mathcal{O}_{2W}}}%
\newcommand{\OTB}{\ensuremath{\mathcal{O}_{2B}}}%
\newcommand{\Oyt}{\ensuremath{\mathcal{O}_{y_t}}}%
\newcommand{\Oyb}{\ensuremath{\mathcal{O}_{y_b}}}%
\newcommand{\cH}{\ensuremath{c_{H}}}%
\newcommand{\cW}{\ensuremath{c_{W}}}%
\newcommand{\cB}{\ensuremath{c_{B}}}%
\newcommand{\cTW}{\ensuremath{c_{2W}}}%
\newcommand{\cTB}{\ensuremath{c_{2B}}}%
\newcommand{\sinthw}{\sin\theta_{\PW}\xspace}%
\newcommand{\sinthwSq}{\swsq}%
\newcommand{\costhw}{\cos\theta_{\PW}\xspace}%
\newcommand{\costhwSq}{\cwsq}%
\newcommand{\tanthw}{\tan\theta_{\PW}\xspace}%
\newcommand{\DgZ}{\ensuremath{\Delta g_1^Z}}%
\newcommand{\DkZ}{\ensuremath{\Delta\kappa_Z}}%
\newcommand{\LPDF}{\ensuremath{\mathcal{W}_{PDF}}\xspace}%
\newcommand{\CS}{\ensuremath{\cos\theta_\mathrm{CS}}\xspace}%
\newcommand{\mee}{\ensuremath{m_{\Pe\Pe}}}%
\newcommand{\mmumu}{\ensuremath{m_{\PGm\PGm}}}%

\cmsNoteHeader{EXO-25-001}
\title{Search for Higgs boson compositeness effects in high-mass dilepton final states in proton-proton collisions at \texorpdfstring{$\sqrt{s} = 13$}{sqrt(s) = 13} and \texorpdfstring{$13.6\TeV$}{13.6 TeV}}

\date{\today}

\abstract{
A search for beyond the standard model effects in high-mass dielectron and dimuon final states with the CMS detector is performed, with a specific focus on Higgs boson compositeness. The search uses data collected at the LHC at 13 and 13.6\TeV center-of-mass energies, corresponding to a total integrated luminosity of 207\fbinv. Compositeness would manifest as distortions in the observed dilepton mass spectrum and the presence of bound states, such as the Higgs boson and new massive vector boson resonances. Assuming the existence of new boson states with masses well above the \TeVns scale, an effective field theory approach provides stringent bounds on the oblique parameters $\Wpar$ and $\Ypar$, which are zero in the standard model. One-dimensional fits, in which only one of the oblique parameters is allowed to deviate from zero, provide the following limits: $-3.9 \times 10^{-4} < \Ypar < 3.5 \times 10^{-4}$ and $-2.1 \times 10^{-4} < \Wpar < 1.6 \times 10^{-4}$ at $95\%$ confidence level. More stringent compositeness limits at the \TeVns scale are set in several new physics scenarios under the assumption that a new \Zprime resonance is physically present within or slightly above the visible dilepton mass spectrum. In this context, cases with large resonance decay widths are explored for the first time.
}

\hypersetup{%
pdfauthor={CMS Collaboration},%
pdftitle={Search for Higgs boson compositeness effects in high-mass dilepton final states in proton-proton collisions at sqrt(s) = 13 and 13.6 TeV},%
pdfsubject={CMS},%
pdfkeywords={CMS, exotica, BSM, composite Higgs}}

\maketitle 
\section{Introduction \label{sec:intro}} 

One of the theoretical conundrums of the standard model (SM) of particle
physics~\cite{Glashow,Weinberg,Salam} is the relatively low value of the mass of the Higgs
boson. Radiative corrections to its mass receive contributions that diverge quadratically
with $\Lambda$, where $\Lambda$ is the highest energy scale for which the SM 
holds~\cite{Veltman}. The fact that the Higgs boson remains light despite these
potentially large corrections is therefore considered unnatural. This is known as
the ``hierarchy problem'', and most attempts to resolve this unsatisfactory feature predict
the existence of new particles and/or interactions around or slightly above the scale of
the electroweak symmetry breaking of the SM. One possibility is to consider that the light
Higgs boson is a composite particle associated with new strong interactions, such as the
so-called Strong Interacting Light Higgs (SILH) scenario~\cite{SILH}. From a generic point
of view, a composite Higgs boson at the CERN LHC~\cite{LHC} could lead to measurable 
signatures such as modified Higgs boson couplings, or excesses in diboson and dilepton 
Drell--Yan (DY) production at high invariant masses. 

Difermion production is an ideal process for measuring the electroweak oblique parameters
\Spar, \Tpar, \Wpar, and \Ypar. These parameters, which vanish in the SM, quantify generic
beyond the SM (BSM) effects from above the electroweak scale by considering modifications in
the exchange of SM electroweak bosons at low energies~\cite{STWY}. While \Spar and \Tpar
have already been significantly constrained at LEP~\cite{PDG2024}, DY production at the
highest energies is expected to be the most sensitive probe for \Wpar and \Ypar. Achieving
ultimate sensitivity in high mass $\llbar$ and $\lnu$ searches at high LHC luminosities
requires a precise understanding of the underlying proton-proton (pp) collisions $\ppVll$
and $\ppWln$ SM processes, at the percent level or better. In the previous expressions and
what follows, the $\Pell$, $\PGn$, $\Vo$, and $\Wo$ symbols are used generically to
represent the appropriate charged ($\Pe^\pm$, $\PGm^\pm$) and neutral ($\PGne, \PGnGm,
\PAGne, \PAGnGm$) leptons, as well as neutral ($\Zo$, $\gamma$) and charged ($\Wp$, $\Wm$)
electroweak gauge bosons, respectively. A key challenge is achieving a precise prediction
of the signal. In this analysis, the signal is modeled using Monte Carlo simulations
incorporating quantum chromodynamics (QCD) effects at next-to-leading order (NLO), while
next-to-next-to-leading order (NNLO) corrections are applied as scale factors on the 
predicted NLO rate.

The only existing LHC results of the search for compositeness effects in $\lnu$ final
states, using Run~2 data at $\sqrt{s} = 13\TeV$, have been published by the CMS
Collaboration in Ref.~\cite{CMS-Run2-ln}. This non-resonant analysis, which is only 
sensitive to the electroweak oblique parameter \Wpar, provided the constraint $\Wpar =
(-1.2^{+0.5}_{-0.6}) \times 10^{-4}$ at the 68\% confidence level (\CL). 

In this paper we search for BSM effects in high-mass dielectron and
dimuon final states, with a specific focus on Higgs boson compositeness. This search is sensitive
to the $\Wpar$ and $\Ypar$ parameters, and is therefore complementary to studies of $\lnu$
final states~\cite{CMS-Run2-ln}. Data from the LHC at $13$ and $13.6\TeV$ center-of-mass
energy are used, amounting to a total integrated luminosity of $207\fbinv$. The
$\ppVll$ reaction proceeds at Born level via a $\qqbar\to\llbar$ hard scattering
process, which can be affected by BSM effects. At QCD NLO, the subprocesses involved
are $\qqVgvirt$, $\qgVq$, and $\qbargVqbar$. In these reactions, $\Pq$ and $\Pg$
generically denote a quark and a gluon, respectively. To minimize theoretical
uncertainties, the SM DY background is generated at QCD NLO precision. Then, event
probabilities are modified to obtain BSM distributions using scale factors that preserve the
original NLO precision.

The paper is structured as follows: Sections~\ref{sec:cmsdetector}
and~\ref{sec:reweighting} introduce the CMS detector, the identification of physics
objects, the data and simulated samples, and the NLO reweighting procedure used in the
analysis. Sections~\ref{sec:selection} and~\ref{sec:corrections} describe the selection
criteria and the additional experimental corrections applied to the simulations in order to
precisely match the observed behavior in the data. Section~\ref{sec:strategy} explains
the fitting strategy employed to measure the oblique parameters $\Wpar$ and $\Ypar$, and
Section~\ref{sec:results} discusses the physics results obtained. A summary of the results
is provided in Section~\ref{sec:summary}. Tabulated results can be found in the HEPData 
record for this analysis~\cite{HEPData}.

\section{The CMS detector and object reconstruction \label{sec:cmsdetector}}

The CMS apparatus~\cite{CMS:2008xjf,CMS:2023gfb} is a multipurpose, nearly hermetic
detector, designed to trigger on~\cite{CMS:2020cmk,CMS:2016ngn,CMS:2024aqx} and identify
electrons, muons, photons, and (charged and neutral)
hadrons~\cite{CMS-electron,CMS:2018rym,CMS:2014pgm}. Its central feature is a
superconducting solenoid of 6\unit{m} internal diameter, providing a magnetic field of
3.8\unit{T}. Within the solenoid volume are a silicon pixel and strip tracker, a lead
tungstate crystal electromagnetic calorimeter, and a brass and scintillator hadron
calorimeter, each composed of a barrel and two endcap sections. Forward calorimeters extend
the pseudorapidity coverage provided by the barrel and endcap detectors. Muons are
reconstructed using gas-ionization detectors interleaved with the layers of the steel flux
return yoke outside the solenoid. More detailed descriptions of the CMS detector, together
with a definition of the coordinate system used and the relevant kinematic variables, can
be found in Refs.~\cite{CMS:2008xjf,CMS:2023gfb}. 

Events of interest are selected using a two-tiered trigger system. The first level,
composed of custom hardware processors, uses information from the calorimeters and muon
detectors to select events at a rate of around 100\unit{kHz} within a fixed latency of
4\mus~\cite{CMS:2020cmk}. The second level, known as the high-level trigger, consists
of a farm of processors running a version of the full event reconstruction software
optimized for fast processing, and reduces the event rate to a few kHz before data
storage~\cite{CMS:2016ngn,CMS:2024aqx}. 

The primary vertex is taken to be the vertex corresponding to the hardest scattering in the
event, evaluated using tracking information alone, as described in
Ref.~\cite{PrimaryVertex}. Electrons are reconstructed from clusters of energy deposits in
the electromagnetic calorimeter. These are matched to hits in the tracker and the
candidates are required to pass a set of identification and isolation
criteria~\cite{CMS-electron} optimized for high energy values. Muons are reconstructed by
matching tracks from the inner tracker with hits, or segments, in the muon system. The
information from the tracker and muon system is combined in such a way as to ensure good
transverse momentum (\pt) resolution and low sensitivity to energy losses due to muon
bremsstrahlung. The muon \pt is determined using a specific combination of different
algorithms optimized for high momentum. The global event reconstruction
(also called particle-flow event reconstruction~\cite{CMS:2017yfk}) aims to reconstruct and
identify each individual particle in an event, with an optimized combination of all
subdetector information. Hadronic jets are clustered from these reconstructed particles
using the infrared and collinear safe anti-\kt algorithm~\cite{Cacciari:2008gp,
Cacciari:2011ma} with a distance parameter of 0.4. Jet momentum is determined as the
vectorial sum of all particle momenta in the jet. 

\section{Signal simulation, reweighting procedure, data and background samples \label{sec:reweighting}} 

An SILH boson can manifest itself through $\qqbar\llbar$ contact
interactions~\cite{Andrea-main}. These interactions should be observed in LHC data,
provided that the masses of the bound states of the new strong sector are not far from the
\TeVns scale. The relevant effective field theory (EFT) operators associated with
the $SU(2)_L$ and $U(1)_Y$ groups of the SM, which are denoted respectively by ${\cal
L'}_{2W}$ and $\mathcal{L'}_{2B}$ respectively, are given by~\cite{WY_theo, Andrea-main}: 
\begin{equation}
	\mathcal{L'}_{2W} = {-\frac{g^2~\Wpar}{\MW^2}~[\PAQq_{L} \gamma_\PGm \tiq \PQq_{L}]~[\overline{\Pell}_{L} \gamma^\PGm \til \Pell_{L}]}; 
	\mathcal{L'}_{2B} = {-\frac{g^{\prime 2}~\Ypar}{\MW^2}~[\PAQq_{\chi_q} \gamma_\PGm \yyq \PQq_{\chi_q}]~[\overline{\Pell}_{\chi_\Pell} \gamma^\PGm \yyl \Pell_{\chi_\Pell}]}, 
\label{eq:Warsaw2}
\end{equation}
where $\mathrm{q}_{\chi_q}$ and $\Pell_{\chi_\Pell}$ denote quark and
lepton fields of either left or right chirality, \ie $\chiq,\chil \in \{L,R\}$. The coupling
constants $g$ and $g'$ are those associated with the $SU(2)_L$ and $U(1)_Y$ groups of the
SM, $\tf$ are the $SU(2)_L$ generators, which only act on left-handed fermions, and
$\yyf$ is the hypercharge of the fermion ${\Pf}_{\chi_\Pf}$. \Wpar and \Ypar are the oblique
parameters introduced in Ref.~\cite{STWY}, which are zero at tree level in the SM,
and the mass of the $\Wo$ boson, $\MW$ has been arbitrarily introduced to provide the right
dimensionality to the contact terms. From a more physical point of view, the right
BSM scale should be $\Lambda$, with $\Lambda\gg \sqrt{s}$, \ie, ${\cal
O}(\Wpar/\MW^2) = \mathcal{O}(\Ypar/\MW^2) = \mathcal{O}(1/\Lambda^2)$.

In the SILH compositeness scenario, \Wpar and \Ypar can be expressed in terms of \gstar,
the typical size of the coupling of the new interactions at high energy, and \mstar, the
energy scale of the new interactions between the fundamental components of the Higgs 
boson~\cite{SILH}:
\begin{equation}\begin{aligned}
	\Wpar \equiv &\cTW~\frac{g^2~\MW^2}{\gstarr~\mstarr}; 
	\Ypar \equiv & \cTB~\frac{g^{\prime 2}~\MW^2}{\gstarr~\mstarr},
\end{aligned}\end{equation}
\noindent expected that the mass of a new $\Zprime$ vector resonance composed of these new
fundamental components and leading to the contact interactions of Eq.~(\ref{eq:Warsaw2}),
has a magnitude of order $\mstar$. 

The procedure applied in this paper is an extension of the reweighting technique for fixed
quark and lepton chiralities described in Ref.~\cite{Andrea-main}. Instead of creating a
grid of dedicated BSM samples for different $\Wpar$ and $\Ypar$ hypotheses, the reweighting
approach takes advantage of the fact that NLO QCD corrections affect BSM and DY amplitudes
similarly. Thus, SM \POWHEG~\cite{POWHEG-Z} $\ppVll$ samples with fixed chiralities are
simply weighted on an event-by-event basis to obtain samples with non-vanishing values of
$\Wpar$ and $\Ypar$, while maintaining NLO QCD precision. This analysis improves upon the
procedure in Ref.~\cite{Andrea-main} to directly obtain the overall event weight averaged
over chiralities. The probability of each chirality combination of the $\qqVg$, $\qgVq$ or
$\qbargVqbar$ subprocesses, generated at the hard scattering level, must still be
multiplied by its differential angular distribution probability in the center-of-mass frame
of the dilepton system, as given in Ref.~\cite{Chaichian}. The resulting expressions
exhibit simplified dependencies in the limit of very high dilepton masses ($q$) compared
with the Z boson mass and the transverse momentum of the dilepton system
(\pt): 
\begin{equation}
	\rho \underset{(q^2\gg\MZ^2,\pt^2)}{\approx} \frac
{\sum\limits_{\chiq,\chil} \left[\left(1+\delta_{\chiq\chil}\CS\right)
~\left(\delta_{LL}~g^2\ttq\ttl~(1-\frac{\Wpar q^2}{\MW^2})+g'^2\yyq\yyl~(1-\frac{\Ypar q^2}{\MW^2})\right)\right]^2 }
{ \sum\limits_{\chiq,\chil} \left[\left(1+\delta_{\chiq\chil}\CS\right)
~\left(\delta_{LL}~g^2\ttq\ttl+g'^2\yyq\yyl\right)\right]^2 }, \label{eq:rho}
\end{equation}
where $\theta_\mathrm{CS}$ is the Collins-Soper angle~\cite{CollinsSoper} and $\ttf$
are the eigenvalues of the third component of weak isospin ($\ttu=+1/2,
\ttd=-1/2,\ttl=-1/2$). The factor $\delta_{\chiq\chil}$ is $+1$ for equal quark and lepton
chiralities ($\chiq=\chil$) and $-1$ for opposite chiralities. The factor $\delta_{LL}$ is
$+1$ when quark and lepton have left-handed quiralities and zero otherwise. In the
analysis, the exact NLO expression of $\rho$, valid for all $q^2$ values, is
used~\cite{Juan-arXiv}.

The correctness of the implementation was cross-checked using a $\ppVll$ SM event sample
generated using \POWHEG~\cite{POWHEG-Z}, inclusive in chiralities. This SM sample was 
reweighted to different values of $W$ and $Y$ following the procedure described above. 
The reweighted samples were then compared with samples generated for dedicated $W$ and 
$Y$ values using a modified version of the reference POWHEG generator, as discussed in 
Ref.~\cite{Andrea-main}.

Run~2 (2016--2018) and Run~3 (2022--2023) data, amounting to the total integrated
luminosities of 200\fbinv (dielectron sample) and 207\fbinv (dimuon sample), were
analyzed. The dielectron and dimuon luminosities differ slightly due to the more stringent
quality criteria required for certification of the dielectron recorded sample. The SM and
signal samples are derived from the original SM DY $\eebar$, $\mmbar$, and $\tautaubar$
samples via reweighting. Besides the reweighted DY component, final states with leptonic
top decays ($\ttbar$, $\PQt\PW$) constitute the dominant background in both the dielectron
and dimuon channels, while diboson final states (WW, WZ, ZZ) are subdominant. Possible
compositeness effects on these backgrounds are expected to be negligible in the sensitive
region of the search. All signal and background events are generated with NLO precision in
QCD using \POWHEG v2~\cite{POWHEG-Z,POWHEG1,POWHEG2,POWHEG3}, except Run~2 WZ and ZZ
samples, which are generated using \PYTHIA 8.2~\cite{pythia82}. The $\ttbar$ and WW
production cross sections are rescaled according to the total NNLO inclusive cross section
calculations~\cite{top-NNLO,WW-NNLO}, and mass-dependent NNLO/NLO QCD
k-factors~\cite{FEWZ} are also applied to the DY samples for consistency. All simulated
predictions are reweighted to the central member of the
PDF4LHC21~\cite{PDF4LHC15,PDF4LHC21} parton distribution function (PDF) set before further
processing.

All generated samples are interfaced with {\PYTHIA} (version 8.2~\cite{pythia82} for Run~2
samples, version 8.3~\cite{pythia83} for Run~3 samples), using
the CP5 tune~\cite{CP5}. The detector response is simulated using the
\GEANTfour~\cite{geant4} package. The presence of additional interactions in the same or
adjacent bunch crossing observed in data (pileup) is incorporated into simulated events by
including extra pp interactions that have been generated with {\PYTHIA}. Simulated samples
are then reweighted to reproduce the same pileup distribution as measured in data. 

\section{Selection criteria \label{sec:selection}}

The selection criteria used in this analysis are similar to those used in the Run~2
search~\cite{CMS-Run2-ll}. The final selection criteria are described in detail below.

\subsection{Dielectron selection criteria}

Only events with two particle candidates with electromagnetic deposits with transverse
energy above 70\GeV (60\GeV in 2016), as measured by the high-level trigger, are used. At
the reconstruction level, the HEEP identification criteria~\cite{CMS-electron} are used to
select the final electrons. These criteria are optimal for high energy electrons. Only
electrons with a transverse energy greater than 100\GeV are used in the analysis.
This requirement effectively suppresses the background from non-prompt electrons
originating from hadronic jets. This background primarily appears in single-electron
background events accompanied by jets from initial state QCD radiation.

The event must contain at least two electron candidates in the kinematic region of
interest, $\mee >200\GeV$. It is required that at least one of the
electrons be in the barrel region ($\abs{\eta}<1.442$). If more than two electrons are present,
the event is rejected if two of the electrons are consistent with the decay of a $\Zo$
boson, $\abs{\mee-91.2} < 20\GeV$. If an event contains two or more pairs of electron
candidates, the pair with the largest $\mee$ is chosen for analysis. 

\subsection{Dimuon selection criteria}

Only events with at least one muon with \pt of at least 50\GeV at trigger level are
included in the analysis. Additionally, alternative trigger paths with different
identification criteria to minimize inefficiencies at high energies, primarily due to
bremsstrahlung effects, are used. For the final selection, we require reconstructed muons
to satisfy specific high-\pt criteria~\cite{CMS-muon}, with $\pt>53\GeV$ and a tracker
isolation requirement of $\sum p_\text{T,tracks}/\pt < 0.1$ in a cone of radius $\Delta
R\equiv\sqrt{(\Delta\phi)^2+(\Delta\eta)^2}<0.3$ around the muon. Here $\sum p_\text{
T,tracks}$ denotes the sum of the \pt of additional tracks in the cone.

Events with dimuon combinations with invariant mass, $\mmumu$, consistent with a Z boson
decay ($\abs{\mmumu-91.2}<20\GeV$), or with a poor quality of the primary vertex (built using
less than four tracks), are rejected. Finally, only events with at least two muon
candidates of opposite electric charge, sufficiently high invariant mass ($\mmumu>200\GeV$)
and with an acollinearity $\pi-\theta_{12}>0.02$, where $\theta_{12}$ is the angular
separation between the two muons, are accepted. The acollinearity requirement is applied in
order to reject potential cosmic ray muon events passing near the primary vertex. If there
are two or more dimuon candidates in the event, the one with the largest $\mmumu$ is
selected for the analysis.

\subsection{Background rejection} \label{sec:back_rej}

Drell--Yan production is treated in a special way in the analysis, since it interferes with
the compositeness signal. The effective dominant backgrounds are thus final states
containing two leptons from processes involving the top quark. Their contribution is
reduced by rejecting selected events with two or more hadronic jets, one of which is
identified as originating from a bottom quark using the DeepJet
algorithm~\cite{Btag-paper,DeepJet}. The jets must have a transverse momentum $\pt > 25\GeV$, a pseudorapidity $\abs{\eta} < 2.5$, and a low probability of being misidentified as a
lepton, with less than 80\% of their energy of electromagnetic or muonic origin. The
applied b~tagging criteria ensure a light-jet contamination at the 1\%
level~\cite{Btag-paper} and provide an efficiency of around 85\% in the typical \pt
range covered by this analysis. Weak diboson production constitutes a minor background to
the selected DY sample. It is estimated using simulated events.

\section{Simulation corrections and comparisons with data \label{sec:corrections}}

For all running periods, the simulated events are weighted according to electron and
muon scale factors determined in data using dileptons with an invariant mass consistent
with the $\Zo$ boson mass~\cite{CMS-electron,CMS-muon}. For electrons, these scale factors
account for discrepancies in cluster reconstruction and high-energy identification.
Dedicated studies show that electron trigger criteria are fully efficient in the phase
space region of the analysis. For muons, the scale factors account for differences between
data and simulation in reconstruction efficiencies, tracker isolation criteria, and trigger
requirements. Additional scale factors are applied in simulation for some periods to
correct for trigger losses caused by accidental misassignments of the bunch crossing
time~\cite{CMS:2020cmk}.

\begin{figure}[htbp]
\centering
\includegraphics[width=0.49\linewidth]{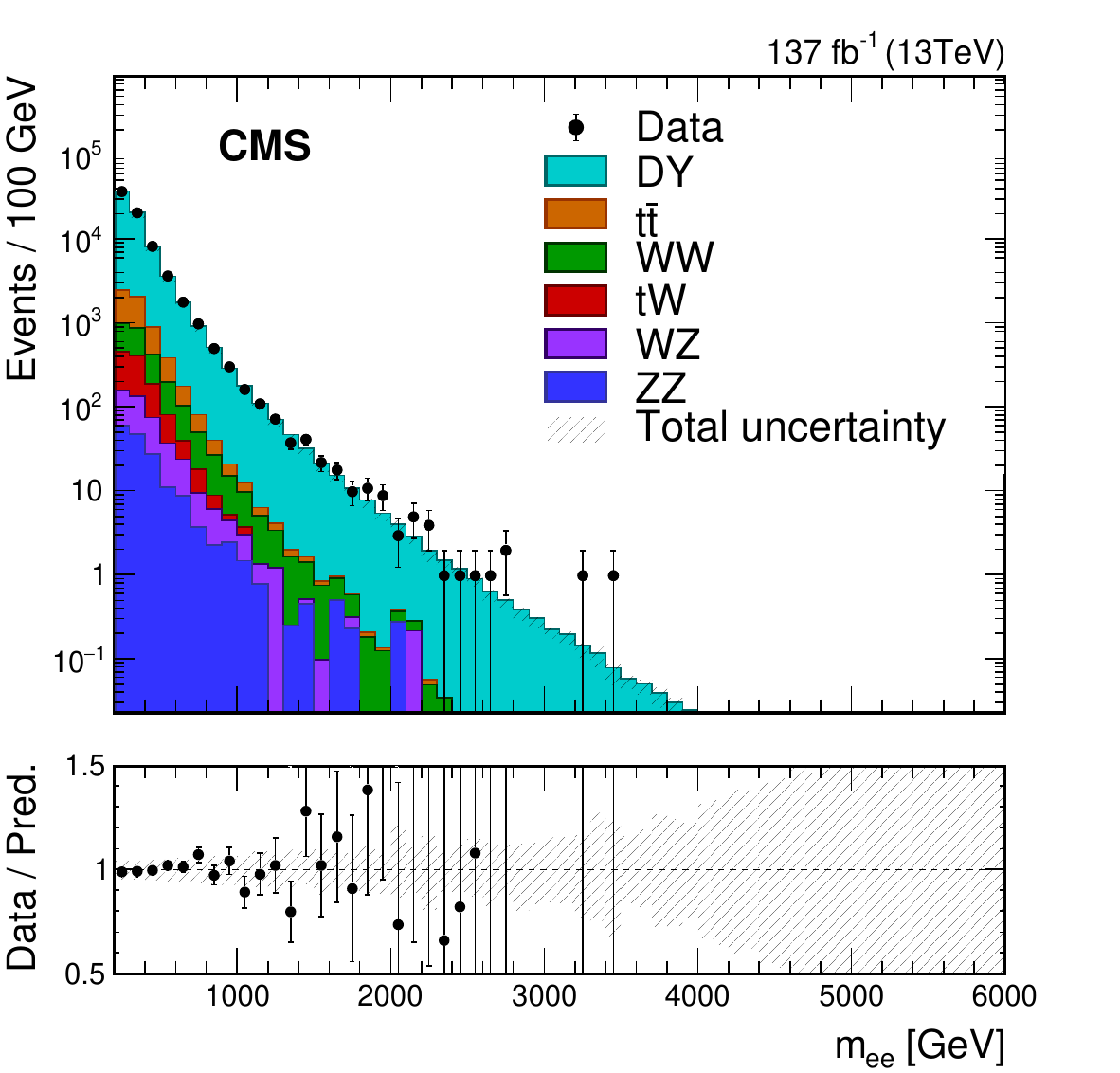}
\includegraphics[width=0.49\linewidth]{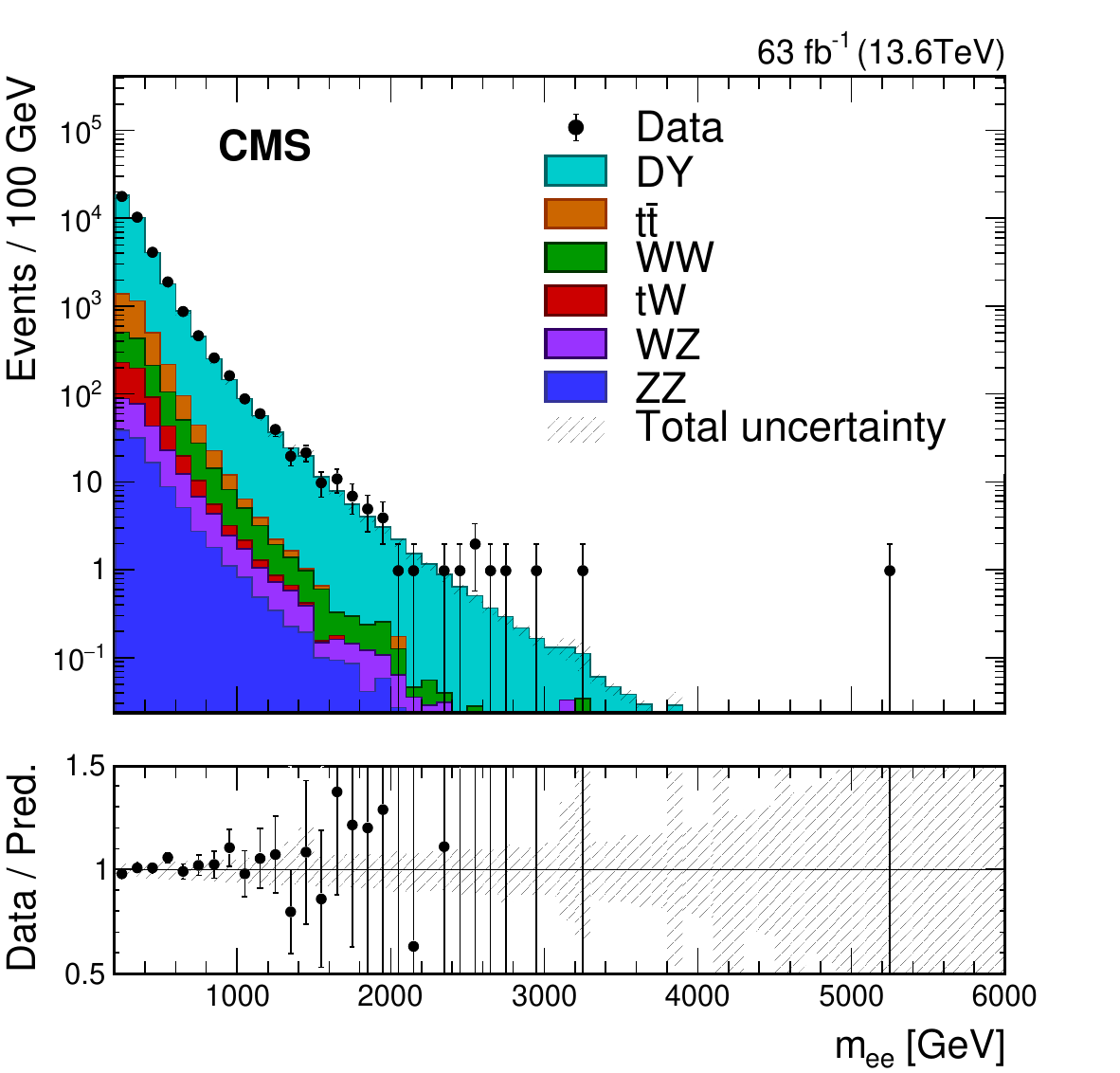}\\
\includegraphics[width=0.49\linewidth]{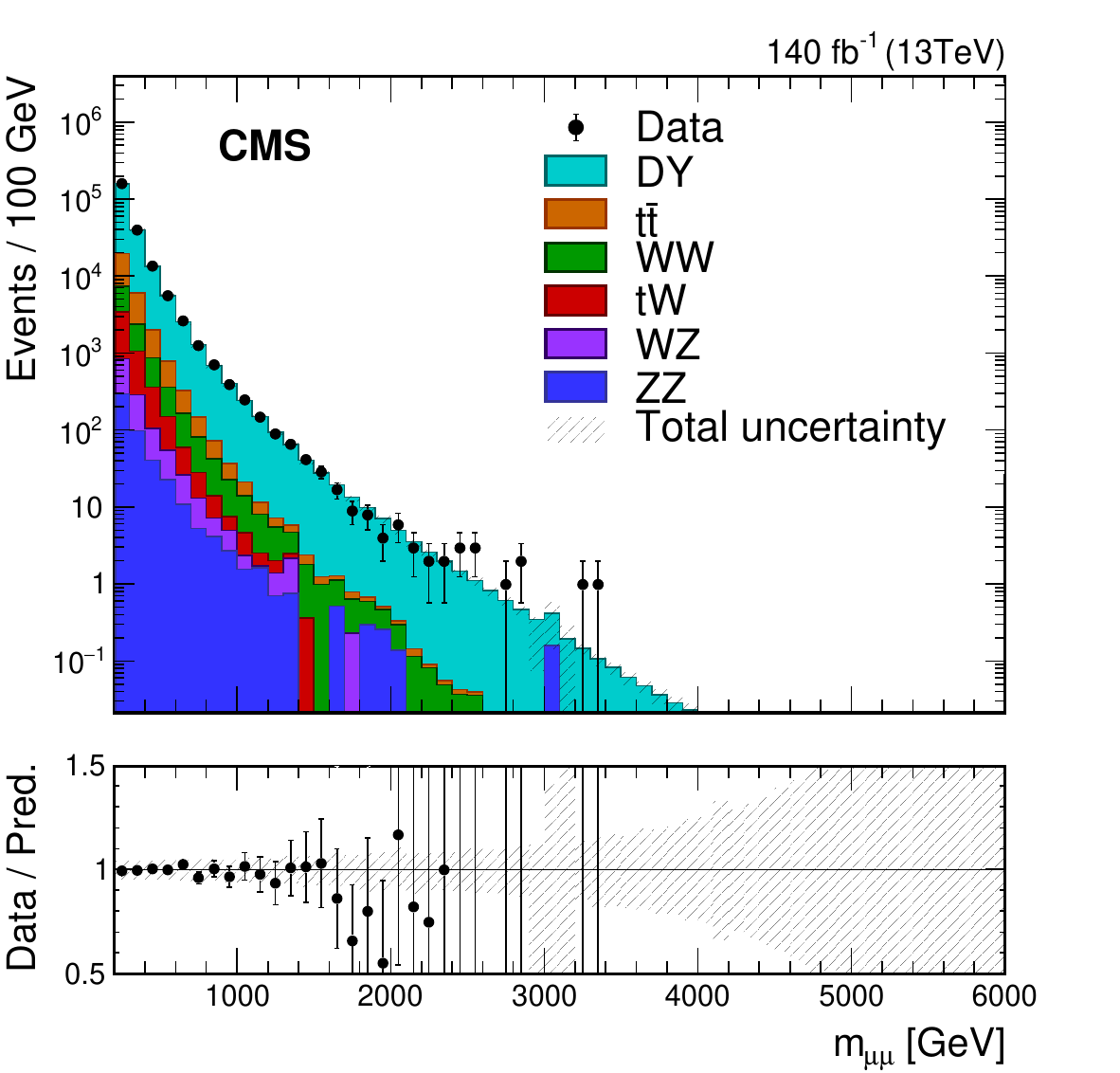}
\includegraphics[width=0.49\linewidth]{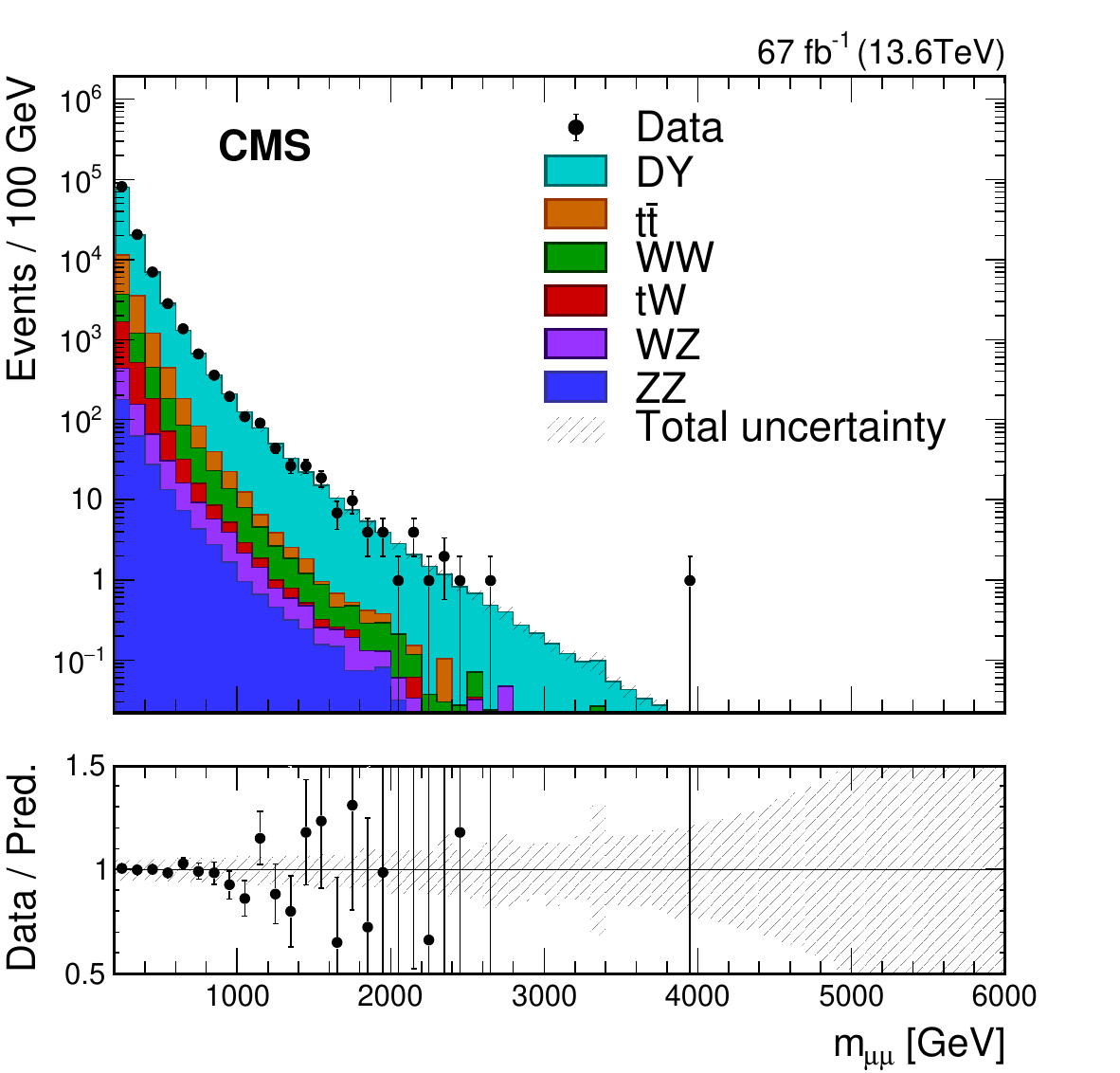}
\caption{Run~2 (left) and Run~3 (right) reconstructed dilepton invariant mass distribution
after selection for (upper) electrons and (lower) muons. The uncertainty bands from the
simulation include statistical uncertainties and systematic effects that modify the overall
differential shape but not the normalization. The ratios between data and predictions 
are also displayed.}
\label{fig:comparisons}
\end{figure}

Electron energy and muon momentum resolutions in the simulation are adjusted according to
the observed resolutions in dilepton data control samples as a function of the running
period. We perform comparisons between data and simulation assuming the PDF4LHC21 PDF set
as reference. The reconstructed $\mee$ ($\mmumu$) distributions for electrons (muons) in
Run~2 and Run~3 after the application of selection criteria are shown in
Fig.~\ref{fig:comparisons}. The $\tautaubar$ contributions are treated as part of the DY
production and are also reweighted in the analysis. Depending on the selection criteria and
the running period, the $\tautaubar$ component constitutes approximately $0.5\%$ to $2\%$
of the total DY selected sample.

This analysis aims to achieve a precision corresponding to NLO in QCD. Using the
PDF4LHC21 PDF set as reference, SM scale factors (``k-factors'') as a function of the
product of the dilepton mass and the sign of \CS have been estimated using
FEWZ~v3{.}1~\cite{FEWZ}. These k-factors take into account NNLO QCD and NLO electroweak (EWK)
corrections, and have variations of order $10\%$ in the 300--6000\GeV mass interval. The DY
SM simulated distributions are corrected accordingly. Additional contributions from
$\PGg\PGg$ initial states, which are relevant at very high dilepton invariant masses,
are also included in the predictions via weighting factors. These contributions are
determined with FEWZ from the ratio of differential cross section calculations that
precisely include photons at the partonic level (PDF4LH15+LUXQED)~\cite{LUXQED-0} to those
that do not include them (PDF4LHC15). 

\section{Analysis strategy, fitting procedure, and systematic uncertainties \label{sec:strategy}}

\begin{figure}[htbp]
\centering
\includegraphics[width=0.9\linewidth]{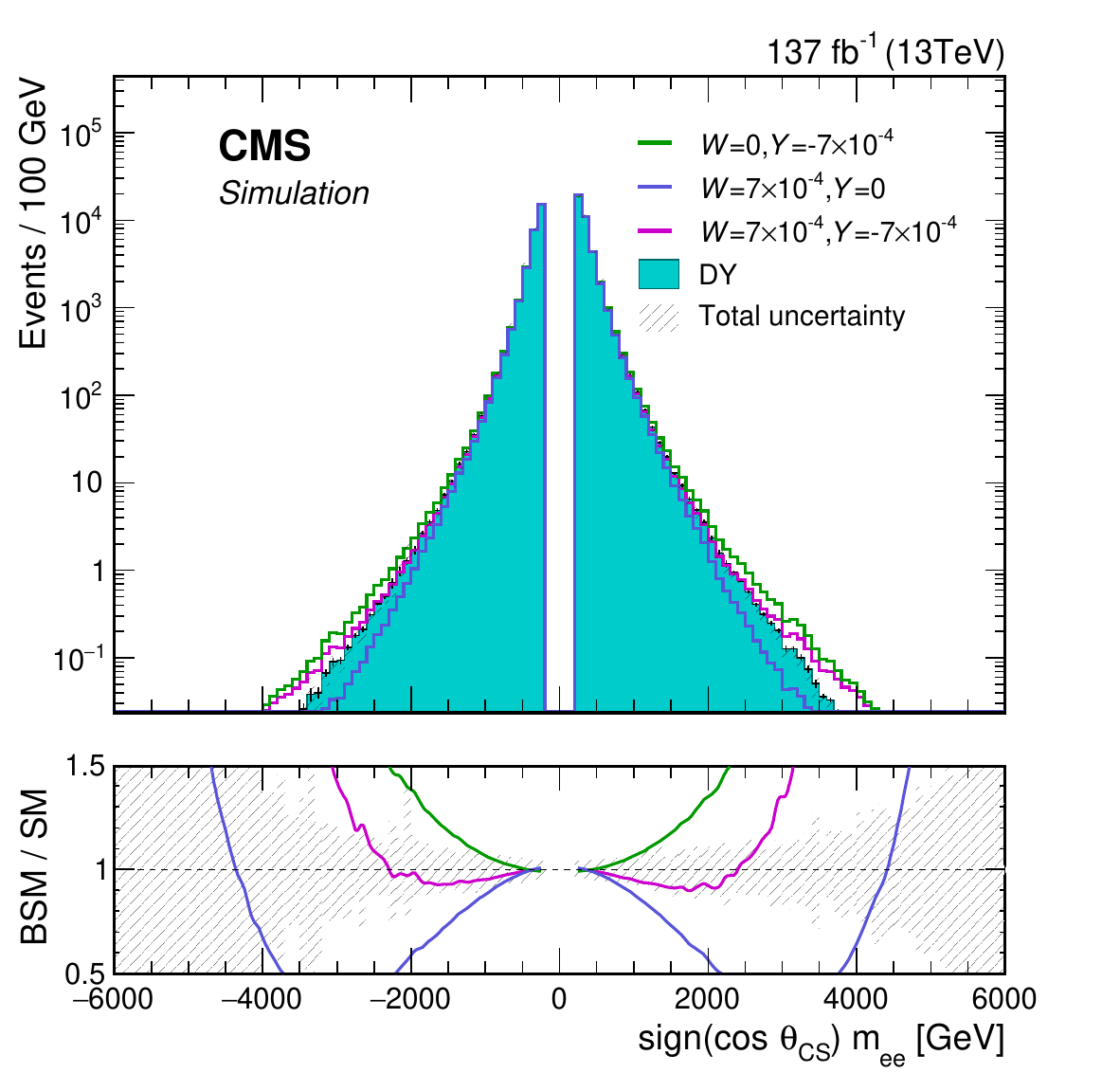}
\caption{Distribution of the product of the simulated invariant mass and the sign of
\CS for DY dielectron production in Run~2, after the application of
selection criteria. The expectations in the SM and in several BSM scenarios, corresponding
to different values of $\Wpar$ and $\Ypar$, are shown. The ratios between data and
predictions are also displayed. The uncertainty band from the simulation includes the
estimated statistical uncertainties and systematic effects.}
\label{fig:BSM2ENT}
\end{figure}

As mentioned in Section~\ref{sec:reweighting}, signal and background simulated samples are
reweighted to the central member of the PDF4LHC21 PDF set~\cite{PDF4LHC21} before the final
analysis. The observed data are then fitted as a function of the compositeness parameters
$\Wpar, \Ypar$. The sensitive distribution is the reconstructed invariant mass of the
selected dilepton events, which is signed according to the reconstructed Collins--Soper
angle~\cite{CollinsSoper}. To define this angle, a direction must be assigned to the
incoming quark in an assumed $\qqVgvirt$ collision. At the reconstruction level, the
direction is approximated as the one that best aligns with that of the dilepton system.
Despite the large uncertainties in the forward-backward asymmetry at high masses, studies
using simulated data show that incorporating this angular information into the analysis
improves the precision of the measured parameters by approximately 5\%. A binned likelihood
fit is performed in the range $(-6000,-300) \cup (300,6000)\GeV$ in bins of
$100\GeV$, with overflows added to the lowest or highest bins, depending on the sign. The
6\TeV boundary safely covers the expected range of CMS sensitivity given the collected
luminosity. The chosen range and bin size were tested to ensure optimal sensitivity for the
expected statistical and systematic uncertainties. The fitting program
{\textsc{Combine}}~\cite{Combine} is used. The DY samples are reweighted as a function
of the compositeness parameters $\Wpar, \Ypar$ at each step of the likelihood minimization
procedure, according to the expressions presented in Section~\ref{sec:reweighting}. While
the CMS $\lnu$ analysis~\cite{CMS-Run2-ln} is only sensitive to $\Wpar$, the $\llbar$ final
state is sensitive to both parameters. Figure~\ref{fig:BSM2ENT} shows how the expected
shape of the signed DY invariant mass distribution for electrons in Run~2 is modified for
different BSM scenarios, corresponding to different values of $\Wpar$ and $\Ypar$. 

The sensitivity is driven by the bin-to-bin distortions of the signed dilepton mass
distribution, with no dependence on the absolute event yield. This is achieved by
introducing a rate parameter that is common to all simulated components (both signal and
background), and which is left completely free in the fit. Constraints on this rate (\eg,
luminosity and overall acceptance within a given year and for each dilepton sample) are
ignored. In this way, the total number of simulated entries is expected to match the total
number of data entries within the statistical uncertainties and the shape systematic
uncertainties. 

Several systematic uncertainties, potentially affecting the shapes of the invariant mass
distributions in a sizable way, are considered in the fit. Their nature and typical input 
sizes in the fits are shown in Table~\ref{tab:systematics}.

\begin{table*}[htbp]
\centering
\topcaption{Systematic uncertainty sources considered in the analysis that have a
non-negligible impact on the shape of the reconstructed invariant mass distribution in the
$0.3 < M < 6\TeV$ mass interval. Their typical sizes, used as input for the likelihood fits, 
are also shown.}
\begin{tabular}{lcl}
 Source & & Typical input uncertainty size \\
\hline
 \rule{0pt}{3ex}PDF uncertainties (including $\alpS$) & & 25--100\% mass shape variations\\
 NNLO QCD + NLO EWK PDF corrections & & ${\approx}$ 10\% mass shape variations\\
 Factorization and renormalization scale & & ${\approx}$ 1--3\% mass shape variations\\
 Final state radiation & & ${\approx}$ 40\% on radiation rate\\
 Initial state radiation & & ${\approx}$ 10\% on radiation rate\\
 Pileup & & ${\pm}$ 4.6\% on total inelastic pp cross section \\
 Top quark background & & ${\approx}$ 1\% (from control sample studies) \\
 b tagging for top quark veto & & ${\approx}$ 3\% on top quark contribution \\
 Diboson background & & 10\% \\
 Muon momentum scale & & $\Delta(1/\mmumu)\approx 1\%~[\TeVns^{-1}]$ \\ 
 Electron energy scale & & 1--2\% on the reconstructed mass \\
 Muon momentum resolution  & & 10\% of nominal resolution \\
\end{tabular}
\label{tab:systematics}
\end{table*}

Uncertainties associated with the PDF choice have the most significant impact on the results
of the analysis. They increase with the dilepton invariant mass, reaching approximately
100\% at the highest masses. Scale uncertainties are considered by doubling or halving the
factorization and renormalization scale values used to evaluate the cross section. The
effects of initial and final state radiation are evaluated by varying the renormalization
scale for emissions in initial and final state radiation. Despite their large size, these
uncertainties depend minimally on the dilepton mass, and their final effect on the fit is
relatively small. Statistical uncertainties from the simulation are small, since the
simulated signals provide more than 10 times the collected amount of data in the relevant
regions of the analysis. This also applies to the DY sample at high masses, where the
reweighting procedure has the greatest impact. The relative uncertainties in the scale
factors used to correct the efficiencies of electrons and muons are typically of
sub-percent size. These uncertainties have a negligible effect on the shape because they
are essentially uncorrelated between different ($\pt,\eta$) bins. Pileup uncertainties are
estimated by varying the total inelastic proton-proton cross section by $4.6\%$ in the
simulation~\cite{pileup}. Given the small size of the measured effect in the data,
uncertainties due to trigger losses caused by accidental misassignments of the bunch
crossing are also considered to be negligible.

Top quark backgrounds are studied in a control sample selected by reverting the b~jet veto
described in Section~\ref{sec:back_rej}. These studies show agreement in shape between
predictions and observations as a function of the dilepton invariant mass. Uncertainties in
the relative normalization between the remaining top quark background and the DY component
are dominated by the uncertainties in the b~tagging scale factors used to correct the
predictions. It is conservatively assumed that the b~tagging systematic uncertainties are
correlated across all running periods. Diboson backgrounds are estimated from simulation
after correction for lepton efficiencies, energy-momentum scale and resolution corrections.
A conservative uncertainty of 10\% on their rate is assumed, consistent with the
statistical level of agreement between predictions and control samples of significant
missing transverse energy and acoplanarity $\pi-\Delta\phi$, where $\Delta\phi$ is the
angular distance between the two leptons in the transverse plane. This control sample is
enriched in dilepton events from diboson final states, mostly from the dominant WW
component. 

Muon momentum scale corrections are studied via the generalized end-point
method~\cite{CMS-muon}. No significant deviations from expectations are observed, but 
the statistical precision of the method is propagated as a systematic uncertainty. The net 
effect is an additional mass shift of order ${\pm}1\%$ at high invariant masses.
Electron energy scale uncertainties in Run~2 lead to a $2\%$ mass scale uncertainty 
for barrel-barrel dielectron events and $1\%$ for barrel-endcap 
candidates~\cite{CMS-Run2-ll}. Similar studies for Run~3 are consistent with a~1\%
mass uncertainty for all dielectron candidates, independently of the rapidity regions.
Nevertheless, it is important to note that the fit is able to constrain a global energy 
scale much more precisely (at the few permille level), so that an initial constraint at the 
1--2\% level has little weight on the final result. This is because an energy 
scale shift is a linear effect, while BSM effects have at least 
a quadratic behavior as a function of the invariant mass.

The limited muon momentum resolution at high energies has a significant impact 
on the measured dimuon invariant mass shape. It leads to a systematic uncertainty 
which is taken into account by an additional ${\pm}10\%$ relative smearing on the 
simulated value of the transverse momentum, according to dedicated 
studies~\cite{CMS-muon}. Systematic resolution effects are substantially smaller for 
electrons at very high energy, since the measurement is totally dominated by 
electromagnetic calorimetry. 

The PDF uncertainties are introduced in the fit by considering all 40 uncorrelated
PDF4LHC21 variations as independent shape nuisance parameters, but fully correlated across
years and samples. These must be treated as shape uncertainties, since the possible range
of oblique parameters results in a variety of shapes of the invariant mass distributions. 
The PDF uncertainties resulting from the limited precision of $\alpS$ are
also considered. Radiation uncertainties are expected to be correlated between different
background components. Renormalization and factorization scale effects are, however,
expected to be largely uncorrelated and are treated as independent.

\section{Results \label{sec:results}}

\subsection{EFT treatment \label{sec:EFT}}

This subsection presents a study based on the assumption of a BSM scale above the
visible mass range, \ie, when $\Lambda\gg 6\TeV$, for which an EFT treatment is
applicable. The one-dimensional fits to the Run~2 and 2022--2023 Run~3 data, including
statistical and systematic uncertainties, lead to the confidence intervals shown in
Table~\ref{tab:1Dfits}. No significant deviations with respect to the SM are observed.

\begin{table*}[htbp]
\centering
\topcaption{Confidence intervals of the one-dimensional fits to the mass to extract $\Ypar$ and $\Wpar$. The value of the non-fitted parameter is fixed to be zero during minimization. The expected boundaries at the 95\% \CL under the SM assumption are also given.}
\begin{tabular}{llll} 
 Fit type & Observed 68\% \CL & Observed 95\% \CL & Expected 95\% \CL \\
\hline 
\rule{0pt}{3ex}$\Ypar$ (imposing $\Wpar=0$) & $[-2.4, 1.0]\times 10^{-4}$ & $[-3.9,3.5]\times 10^{-4}$ & $[-3.4,3.9]\times 10^{-4}$ \\
 $\Wpar$ (imposing $\Ypar=0$) &  $[-1.2, 0.7]\times 10^{-4}$ & $[-2.1,1.6]\times 10^{-4}$ & $[-2.0,2.0]\times 10^{-4}$\\
\end{tabular}
\label{tab:1Dfits}
\end{table*}

\begin{figure}[htbp]
\centering
\includegraphics[width=0.8\linewidth]{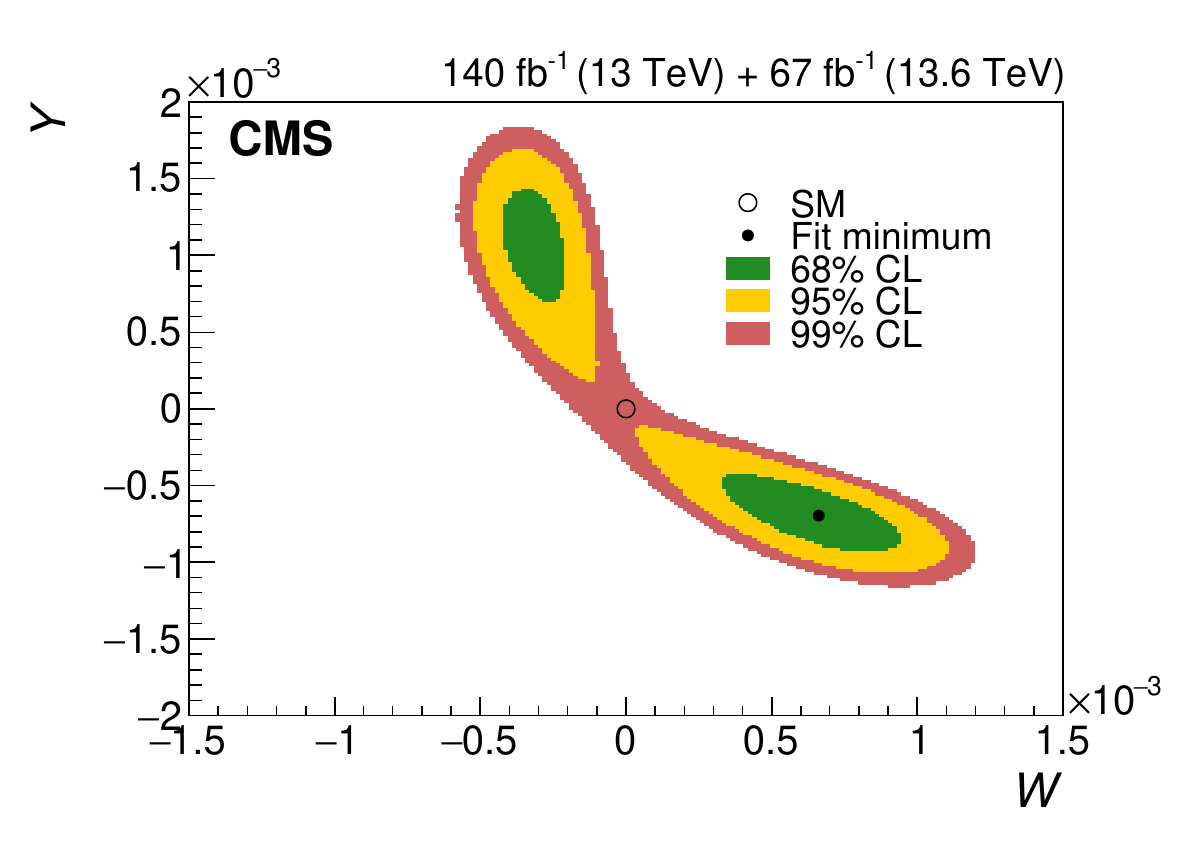}
\caption{Allowed regions in the 2-dimensional $(\Wpar,\Ypar)$ parameter plane, from a fit
to the full data sample considered in the analysis. The fit minimum corresponds to
$(\Wpar,\Ypar)\approx (+6.6\times10^{-4}, -7.0\times10^{-4})$. There is consistency with the
SM at the two-standard deviation level. This deviation is largely driven by the presence of a
dielectron event observed in 2022 data with an invariant mass above 5\TeV.} 
\label{fig:2Dfit}
\end{figure}

Results from a two-dimensional fit as a function of the parameters $\Wpar$ and $\Ypar$ are
shown in Fig.~\ref{fig:2Dfit}. A departure from the SM at the two-standard deviation level,
which corresponds to a 95.7\% \CL, is visible. This is largely driven by the presence of a
2022 dielectron candidate with an invariant mass of $5.2\TeV$. Removing this event from the
sample would result in a departure at the level of one standard deviation. This
two-dimensional fit is less sensitive to SM deviations in regions in the ($\Wpar, \Ypar$)
plane where there are significant cancellations of BSM effects. As shown in
Eq.~(\ref{eq:rho}), there are terms that deviate linearly from the SM proportionally to
$\Wpar$ and $\Ypar$, as well as terms that exhibit a quadratic dependence. Regions with
less sensitivity correspond to a dominance of the quadratic terms, which can only be
constrained at the highest observed masses where the expected number of events is low. This
behavior is typical of dimension-eight effects, which are not specifically targeted by this
analysis. Note that large cancellations of the linear dependencies do not occur in the
one-dimensional fits, which explains their higher constraining power and better agreement
with SM expectations. 

\begin{figure}[htbp]
\centering
\includegraphics[width=0.75\linewidth]{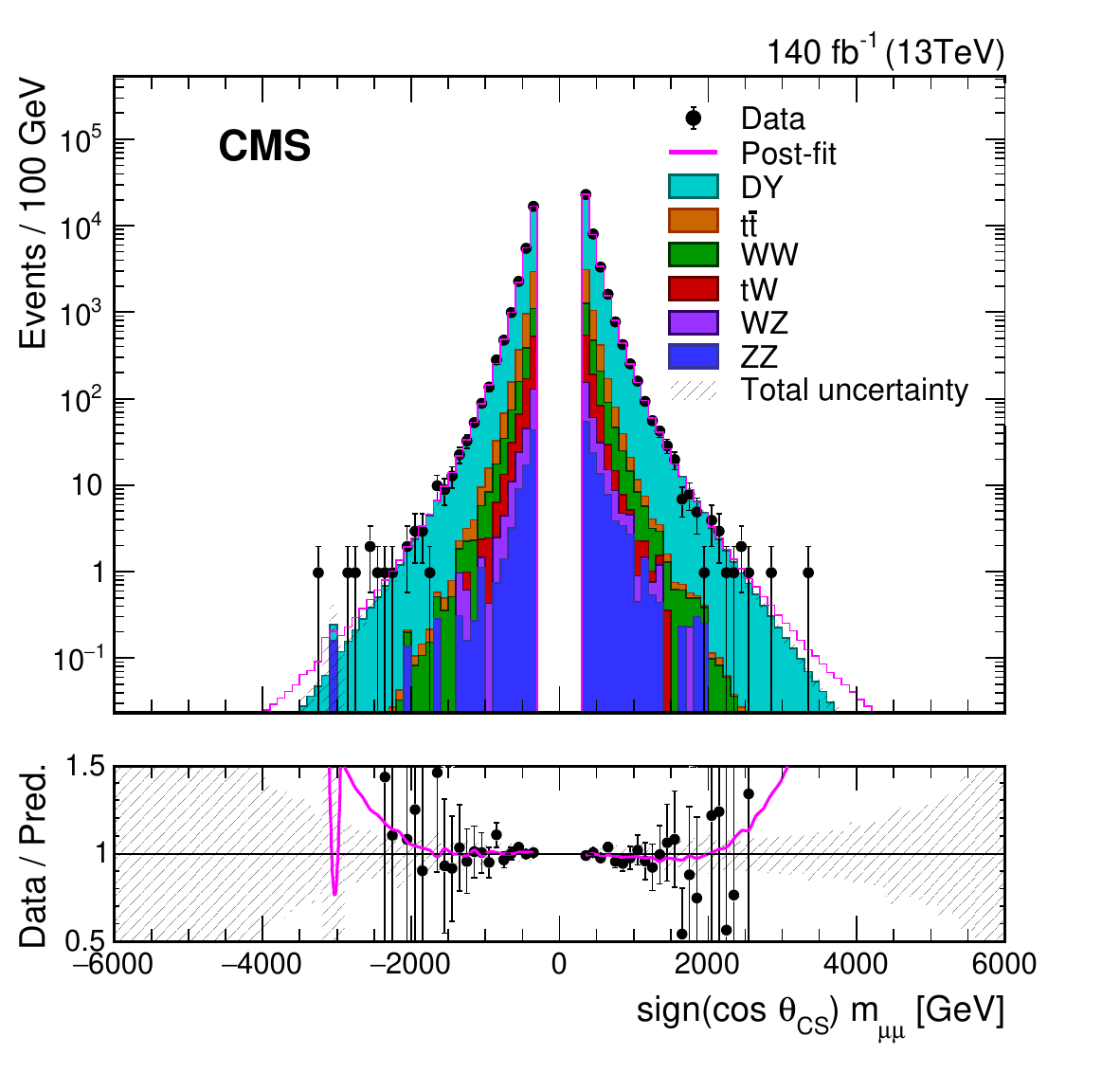}\\
\includegraphics[width=0.75\linewidth]{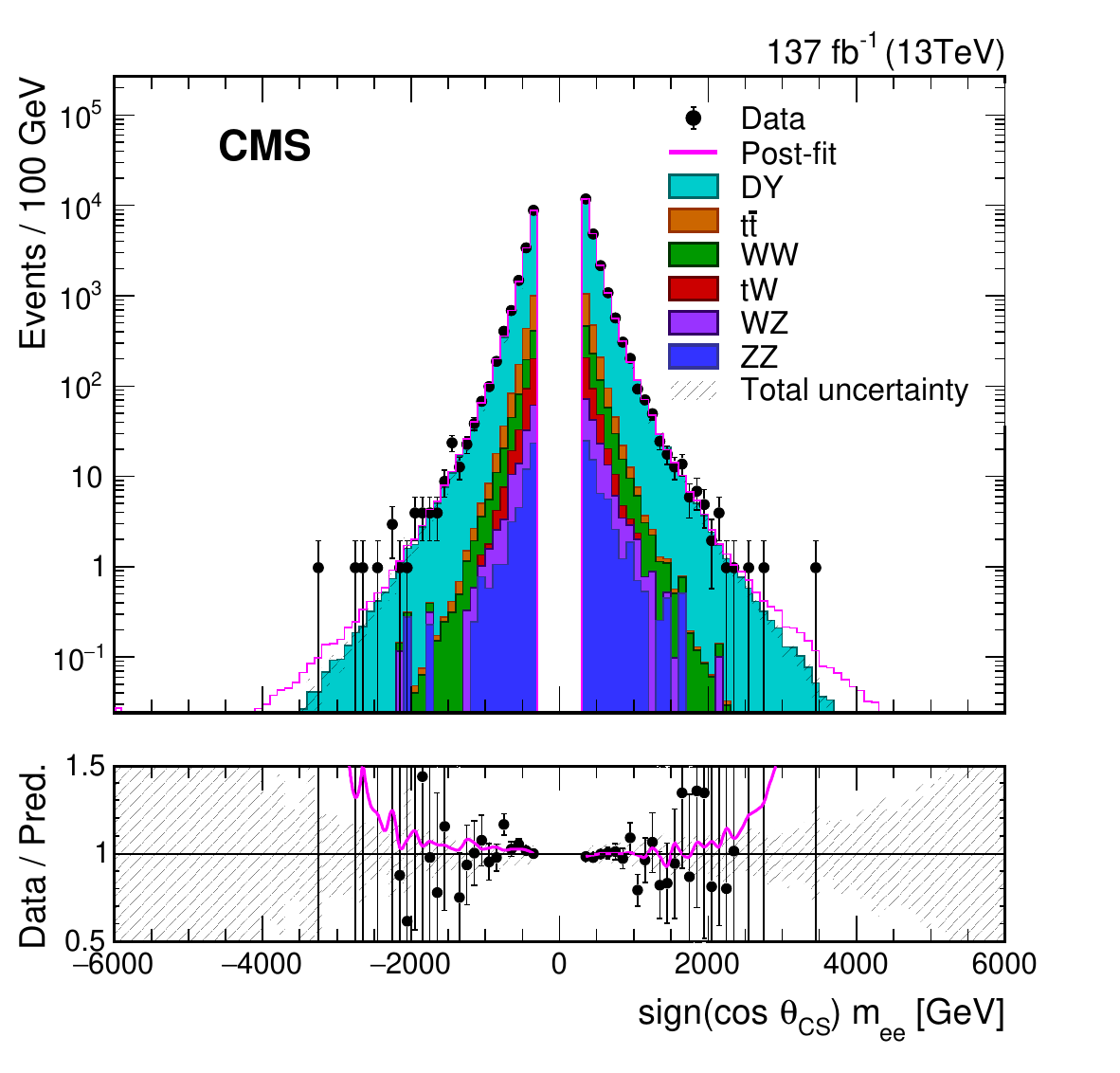}
\caption{Two-dimensional EFT fit results superimposed on the observed data and SM
predictions, for Run~2 dimuons (upper plot) and dielectrons (lower plot). The ratios
between data and predictions are also displayed.} 
\label{fig:postfit-distributions1}
\end{figure}

\begin{figure}[htbp]
\centering
\includegraphics[width=0.75\linewidth]{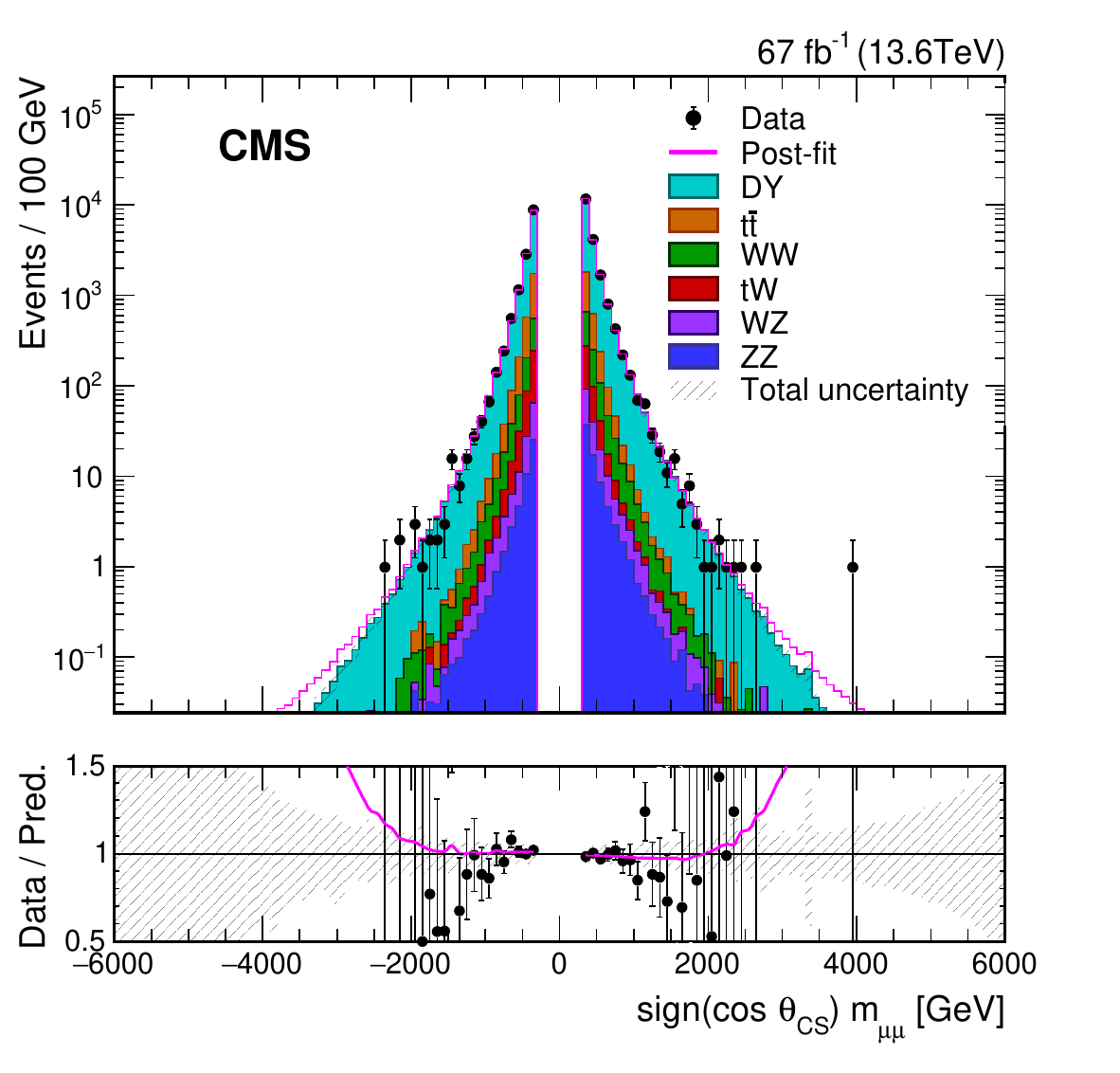}\\
\includegraphics[width=0.75\linewidth]{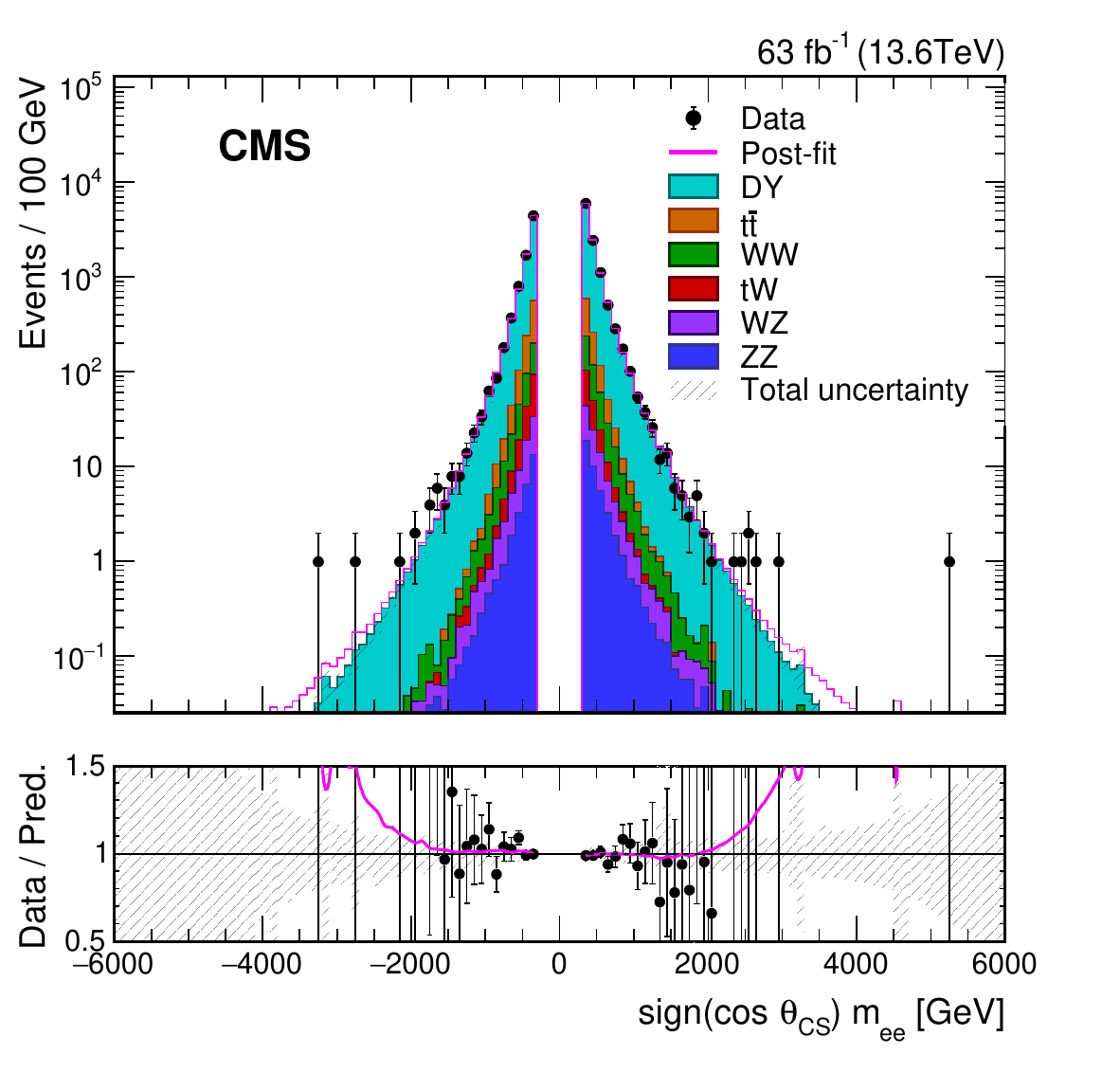}
\caption{Two-dimensional EFT fit results superimposed on the observed data and SM
predictions for Run~3 dimuons (upper plot) and dielectrons (lower plot). The ratios between
data and predictions are also displayed.} 
\label{fig:postfit-distributions2}
\end{figure}

The largest contributions to the systematic uncertainty are associated with theoretical
sources that significantly change the shape of the mass distribution at large values, such
as PDFs, k-factors, and renormalization and factorization scale uncertainties.
Figures~\ref{fig:postfit-distributions1}-\ref{fig:postfit-distributions2} show the
distributions for the different dilepton samples and running periods after fitting. These
distributions are superimposed on the SM predictions with the estimated systematic
uncertainties in the shape.

\begin{figure}[htbp]
\centering
\includegraphics[width=0.8\linewidth]{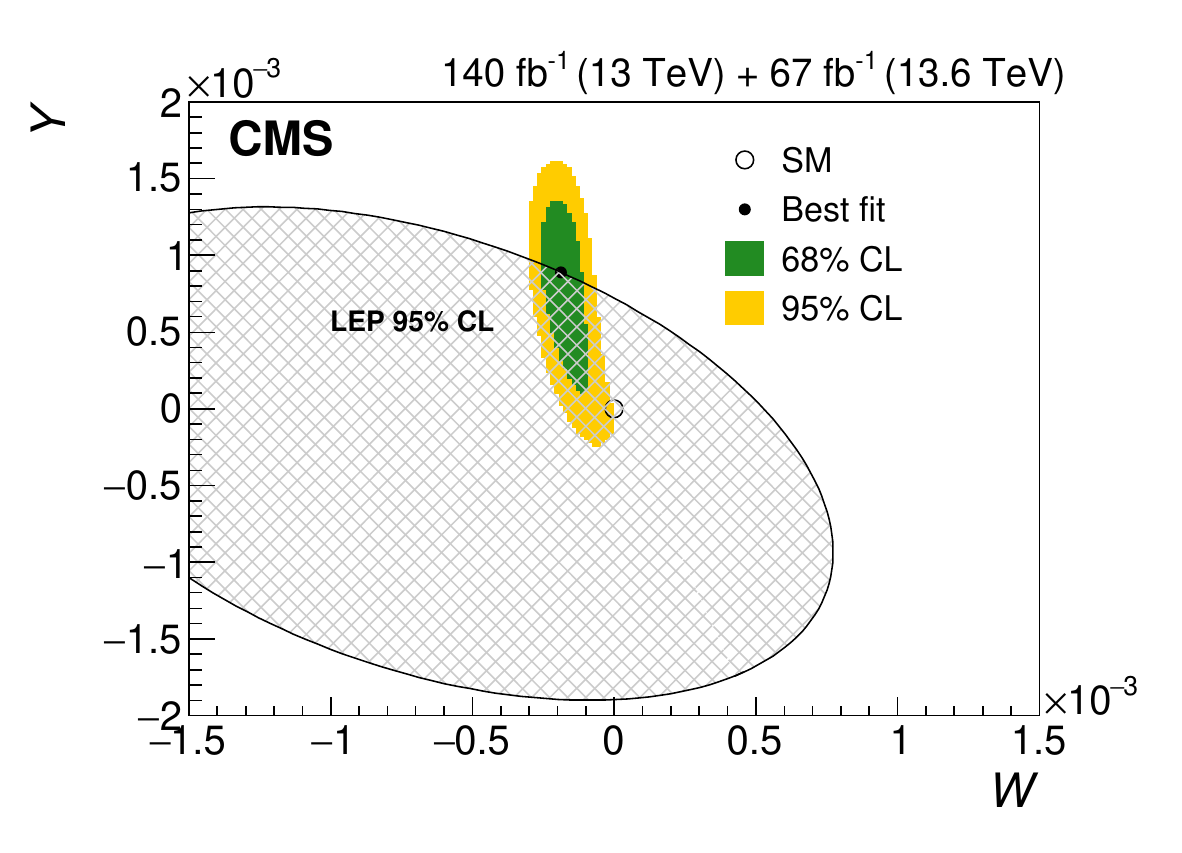}
\caption{Allowed region in the $(\Wpar,\Ypar)$ parameter plane, using the full data sample,
when the $\Wpar$ constraint~\cite{CMS-Run2-ln}, $\Wpar = (-1.2^{+0.5}_{-0.6}) \times
10^{-4}$ is included in the fit. The fit minimum corresponds to $(\Wpar,\Ypar)\approx
(-1.9\times10^{-4}, +8.9\times10^{-4})$. The SM is consistent with the observations at the
two-standard deviation level. The figure also shows the region constrained by LEP
electroweak measurements~\cite{LEP}.} 
\label{fig:WconstrainedFit-data}
\end{figure}

Figure~\ref{fig:WconstrainedFit-data} shows the allowed region for an SM scenario when the
$\Wpar$ constraint from the CMS Run~2 $\lnu$ search~\cite{CMS-Run2-ln}, $\Wpar =
(-1.2^{+0.5}_{-0.6}) \times 10^{-4}$, interpreted as a $\chi^2$ term, is added to the
current two-dimensional fit. Although a deviation from the SM at the two-standard deviation
level is still observed, driven partly by the $\Wpar$ constraint, this new CMS result
strongly constrains any BSM scenario that predicts non-zero values for the oblique
parameters, $\Wpar$ and $\Ypar$. 

\begin{figure}[htbp]
\centering
\includegraphics[width=0.8\linewidth]{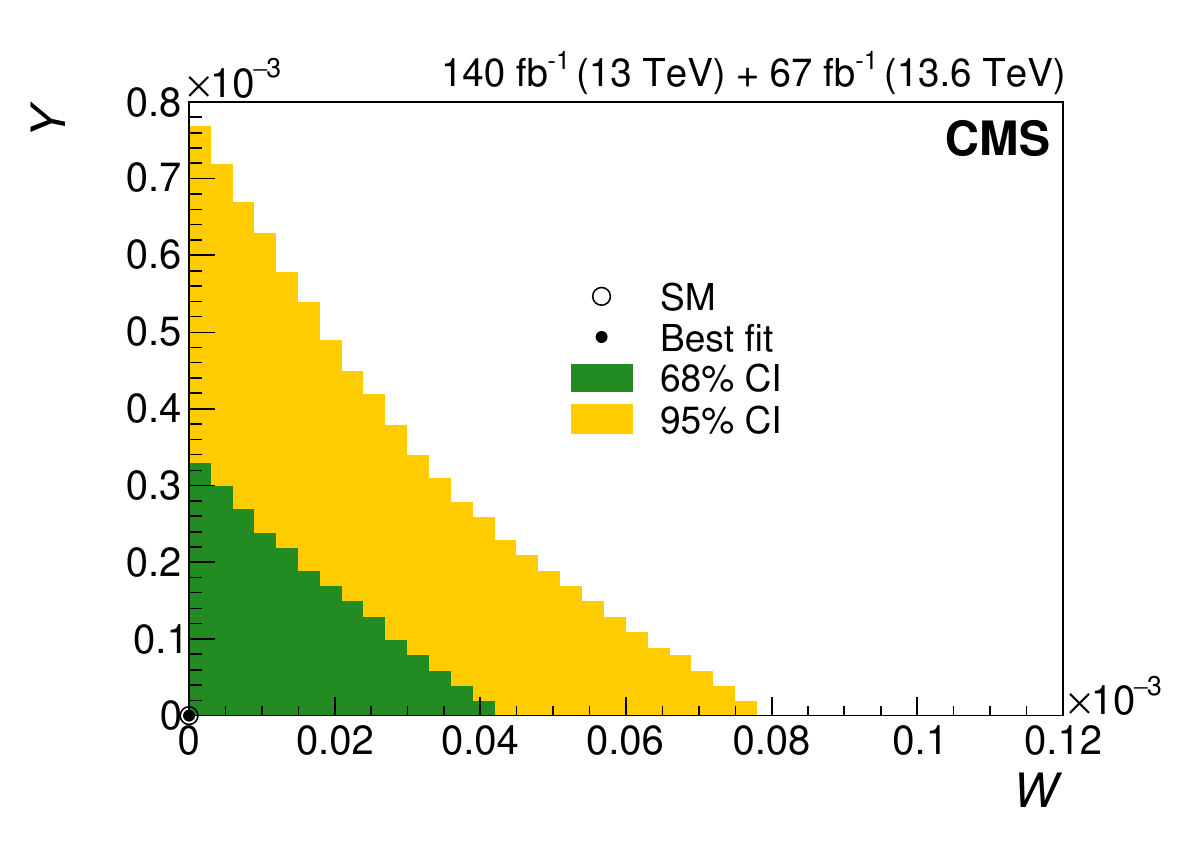}
\caption{Allowed regions as a function of $\Wpar$ and $\Ypar$ under a universality
assumption, which requires $\Wpar\ge 0$ and $\Ypar\ge 0$. The likelihood is
derived from the analysis of the full Run~2+2022+2023 data sample of dielectrons and
dimuons. The green and yellow areas correspond to the allowed 68\% and 95\% Bayesian
credible intervals (CI)~\cite{PDG2024}, under the positivity constraint and a flat prior
assumption for the parameters.} 
\label{fig:2Dfit-universal}
\end{figure}

While the observed discrepancy could result from a real BSM effect, largely driven by the
very high-mass dielectron event present in the sample and the Run~2 $\Wpar$ constraint, the
SILH scenarios discussed in Refs.~\cite{McCullough-etal, SILH} require positive values of
$\Wpar$ and $\Ypar$. Figure~\ref{fig:2Dfit-universal} displays the 68\% and 95\% Bayesian
credible intervals~\cite{PDG2024}, assuming such a positivity constraint and a flat prior
distribution as a function of $\Wpar$ and $\Ypar$, given the approximately linear
dependence of the cross section on these parameters. The 95\% credible region lies within
the intervals $0 < \Wpar < 7.8\times 10^{-5}$ and $0 < \Ypar < 7.6\times 10^{-4}$. 

Table~\ref{tab:universal} collects the 95\% credible bounds on the BSM scale 
associated with several universal scenarios characterized by slightly different values of
the Wilson coefficients, $\cTW$ and $\cTB$. Owing to the presence of a single $\mstar\gstar$
scale, all models lead to a relationship between $\Wpar$ and $\Ypar$ such that only one of
the compositeness parameters is varied in the fit. The flat prior assumption for the 
compositeness parameter is also consistent with a flat prior for the cross section and a 
Gaussian behavior of the likelihood in the explored regions.

\begin{table*}[htbp]
\centering
\topcaption{EFT constraints on the oblique parameters $\Wpar$ and $\Ypar$ for several
universal scenarios: Y-universal~\cite{Y-universal}, Little Higgs~\cite{LittleHiggs,SILH},
Little Higgs with custodial symmetry~\cite{CustodialLittleHiggs,SILH}, and holographic
composite Higgs with equal bulk gauge couplings~\cite{HolographicHiggs,SILH}). The quoted
bounds on $\mstar\gstar$ correspond to the 95\% Bayesian credible intervals~\cite{PDG2024}.} 
\begin{tabular}{lc} 
Universal scenario & Lower bound on $\mstar\gstar [\TeVns{}]$ \\
\hline
\rule{0pt}{3ex}Y-universal ($\cTW=0$, $\cTB=1$) & $\phantom{1}1.2$ \\
Little Higgs ($\cTW=1$, $\cTB=0$) & $\phantom{1}7.0$ \\
Little Higgs with custodial symmetry ($\cTW=1$, $\cTB=1$) & $\phantom{1}7.2$ \\
Holographic composite Higgs ($\cTW=1.9$, $\cTB=3.8$) & $10.1$ \\ 
\end{tabular}
\label{tab:universal}
\end{table*}

\newpage

\subsection{\texorpdfstring{\PZpr}{Z'} resonance treatment}

This subsection evaluates the sensitivity of CMS data to the presence of a nearby $\Zprime$
resonance in a compositeness scenario, either within the visible mass range or slightly
above it. To this end, we assume that, if BSM effects are due to the presence of a
$\Zprime$ resonance of mass $\mstar$, its EFT approximation corresponds to a scenario with
$\Lambda\equiv\mstar$ in the $\sqrt{s} \ll \mstar$ limit. The resonant treatment in the
analysis is simply obtained by introducing in the amplitudes the additional Breit--Wigner
factor $F_{BW}=-\mstarr/(s'-\mstarr+i\mstar\Gamma^*$), where $\sqrt{s'}$ is the dilepton
invariant mass and $\Gamma^*$ is the $\Zprime$ width. In the $\sqrt{s'}\ll \mstar$ limit
one recovers the EFT treatment: $F_{BW}=1$. This approach also allows an interpretation of
the $\Zprime$ resonance couplings in the context of heavy vector triplet (HVT)~\cite{HVT}
and heavy vector singlet (HVS)~\cite{HVS} scenarios via $\Wpar$ and $\Ypar$ parameters. 

\begin{figure}[htbp]
\centering

\includegraphics[width=0.8\linewidth]{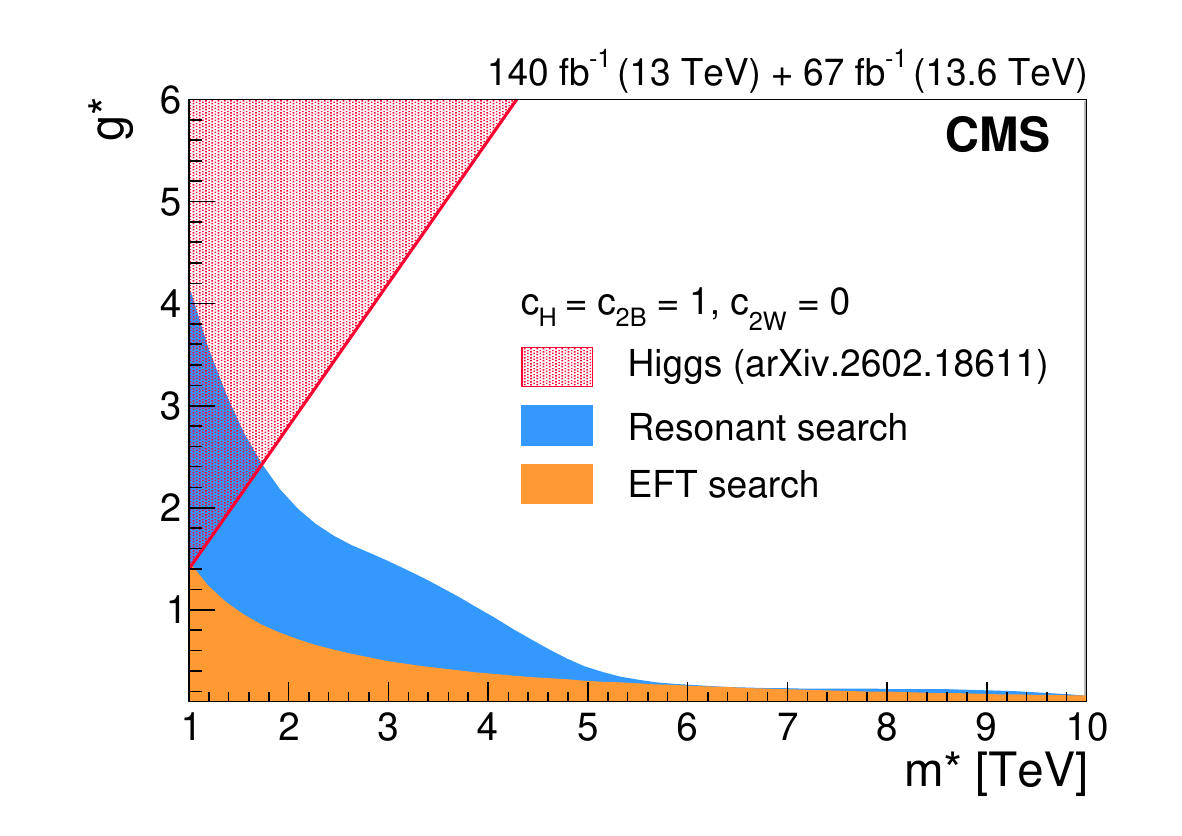}

\caption{Excluded regions at the 95\% \CL in the $(\mstar,\gstar)$ plane in a Y-universal 
scenario~\cite{Y-universal}, assuming
the presence of a neutral gauge boson $\Zprime$ of mass $\mstar$. Decays into $\WW$ and
$\ZH$ final states are assumed to increase the width of the new resonance, which is
explicitly defined as a function of the strong coupling constant $\gstar$ of the SILH model
(see text). These bounds are compared with those deduced from a non-resonant, EFT fit
and from the current CMS global fits of the Higgs boson signal strength modifier for $\cH=1$
($\gstar<1.4~\mstar[\TeVns]$ at 95\% \CL)~\cite{CMS:HIGGSNEW}.}

\label{fig:Yuniversal}

\end{figure}

\begin{figure}[htbp]
\centering
\includegraphics[width=0.8\linewidth]{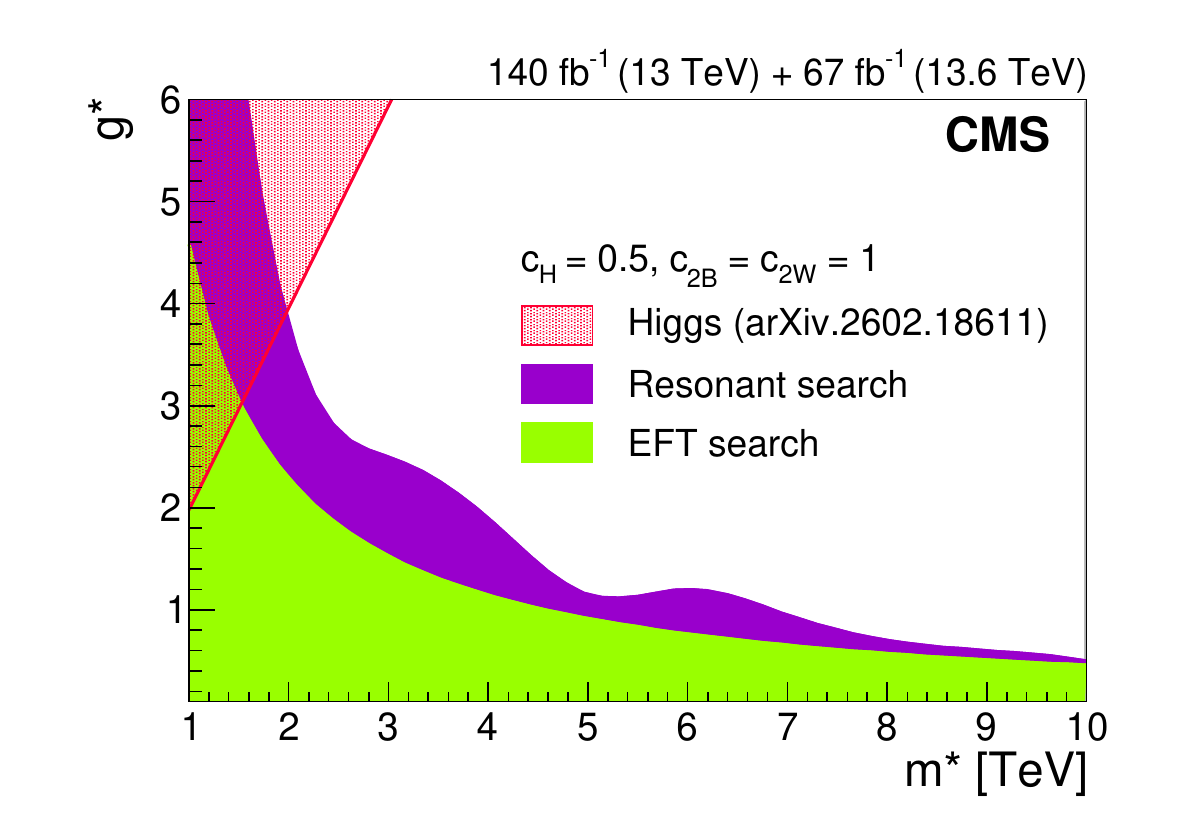}
\caption{Excluded regions at the 95\% \CL in the $(\mstar,\gstar)$ plane in a
Little Higgs scenario with custodial symmetry~\cite{CustodialLittleHiggs}, assuming
the presence of a neutral gauge boson $\Zprime$ of mass $\mstar$. Decays into $\WW$ and
$\ZH$ final states are assumed to increase the width of the new resonance, which is
explicitly defined as a function of the strong coupling constant $\gstar$ of the SILH model
(see text). These bounds are compared with those deduced from a non-resonant, EFT fit
and from the current CMS global fits of the Higgs boson signal strength modifier for $\cH=0.5$
($\gstar<2~\mstar[\TeVns]$ at 95\% \CL)~\cite{CMS:HIGGSNEW}.}
\label{fig:CustodialLittleHiggs}
\end{figure}

First, a Y-universal scenario is considered in which a new neutral gauge boson couples to
SM fermions according to hypercharge. This model remains anomaly-free regardless of the
number of right-handed neutrinos considered~\cite{Y-universal}. The width of the new
resonance is specified as a function of both $\gstar$ and $\mstar$. In this context, we
consider an HVS scenario in which contributions to $\lnu$ final states are absent:
$\Wpar=0, \Ypar = (g'^2~\MW^2)/(\gstarr~\mstarr)$, but decays into $\WW$ and $\ZH$ are
naturally allowed~\cite{HVT}. This choice would imply:
\begin{equation}
\frac{\Gamma^*}{\mstar}  =  \frac{5 g'^4}{12\pi~\gstarr} + \frac{\gstarr}{96\pi}.
\end{equation}
Additionally, a $\Zprime$ resonance within a Little Higgs scenario with custodial symmetry
is considered~\cite{CustodialLittleHiggs,SILH}. In this case, both $\Wpar$ and $\Ypar$
couplings contribute coherently: $\Wpar/g^2=\Ypar/g'^2 = \MW^2/(\gstarr~\mstarr)$, and
the total width of the resonance is given by:
\begin{equation}
\frac{\Gamma^*}{\mstar} = \frac{3 g^4+5 g'^4}{12\pi~\gstarr} + \frac{\gstarr}{96\pi}.
\end{equation}
Figures~\ref{fig:Yuniversal}-\ref{fig:CustodialLittleHiggs} show the constraints imposed by
CMS data in the $(\mstar,\gstar)$ parameter plane in these two scenarios. The obtained
bounds are also compared with those derived from the current CMS global fits of the Higgs
boson signal strength modifier~\cite{CMS:HIGGSNEW}, 
$\gstar< 1.4~\mstar/\sqrt{\cH}~[\TeVns{}]$ at 95\% \CL, and with
the constraints obtained from a non-resonant treatment using the one-dimensional EFT fits
of Y, as described in Section~\ref{sec:EFT}.

\begin{figure}[htbp]
\centering
\includegraphics[width=0.49\linewidth]{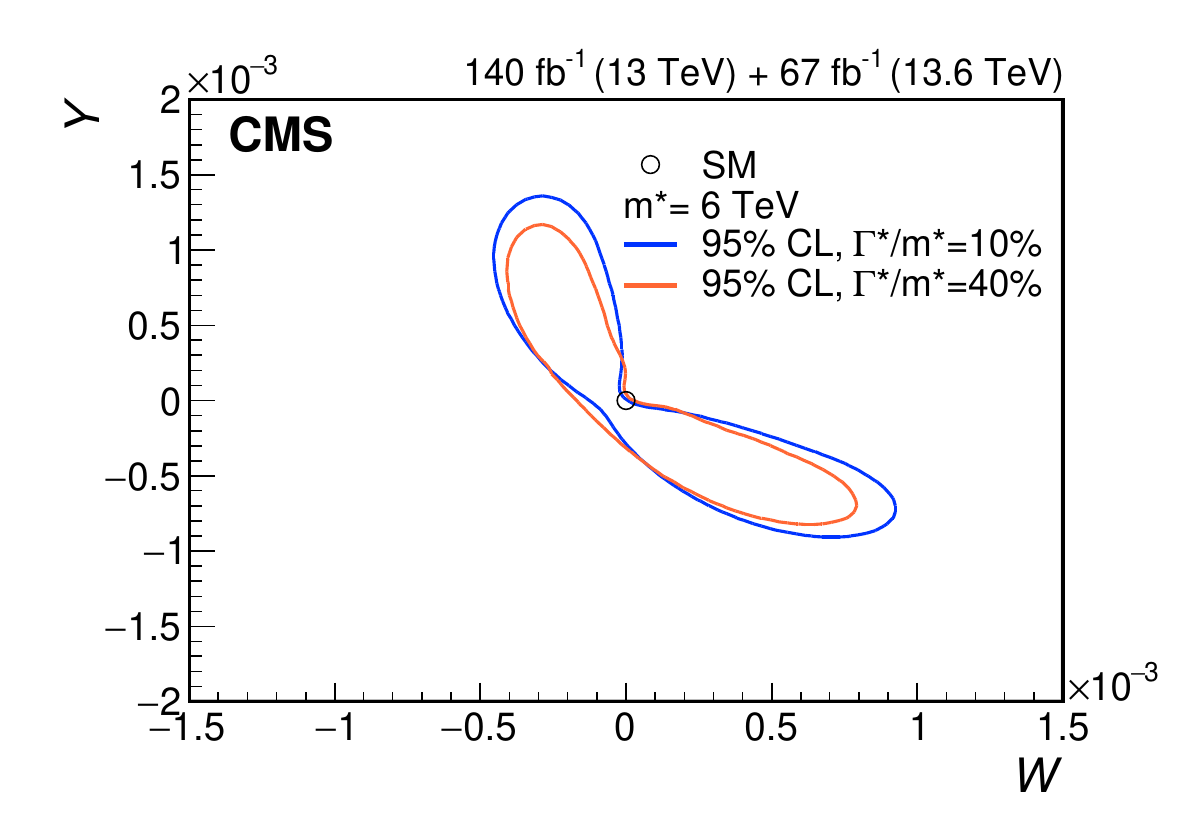}
\includegraphics[width=0.49\linewidth]{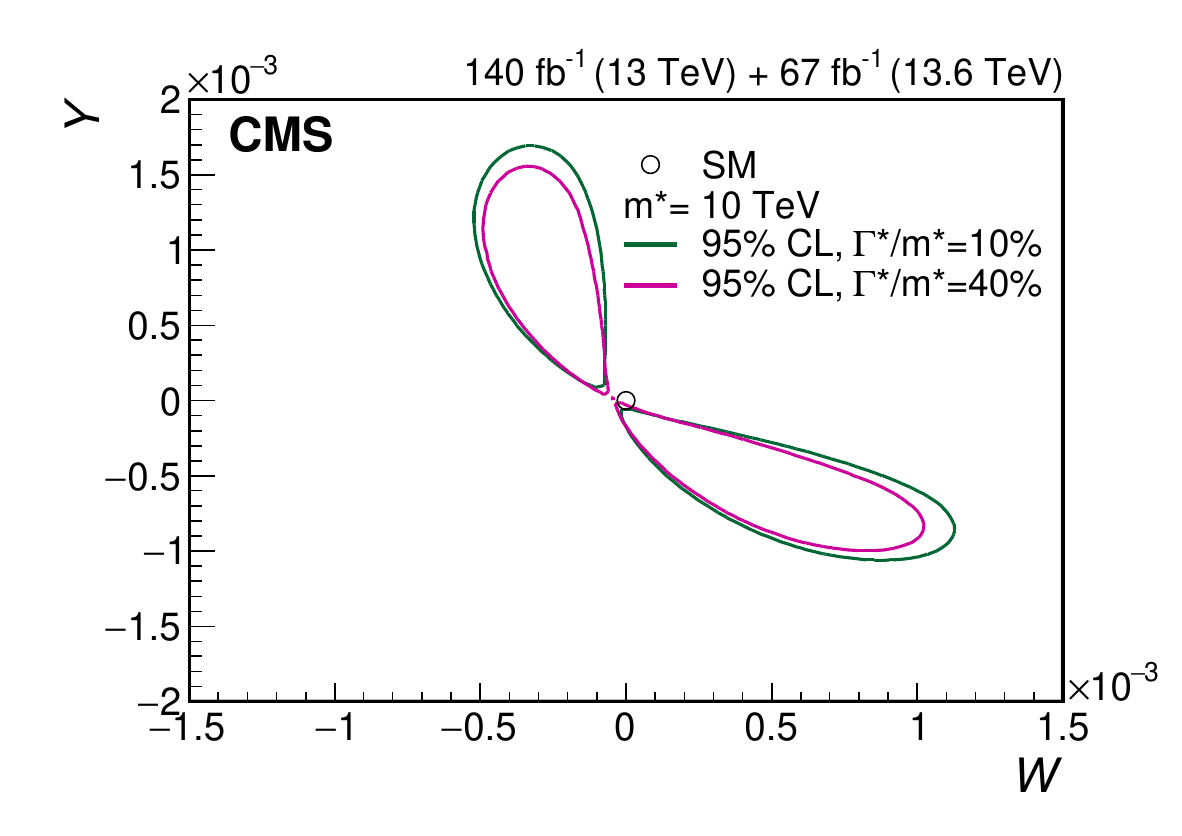}
\caption{Allowed regions at the 95\% \CL in the 2-dimensional $(\Wpar,\Ypar)$ parameter plane, assuming a
nearby BSM energy scale. The presence of neutral gauge boson resonances of $6\TeV$
(left) and $10\TeV$ (right) is assumed, with a medium (10\%) or large (40\%) relative decay
width, due to the potential presence of a significant decay rate into $\WW$ or $\ZH$ final
states.} 
\label{fig:2Dfit-with-resonances}
\end{figure}

Finally, scenarios involving new $\Zprime$ bosons with medium (10\%) and large (40\%)
relative widths are explored. These scenarios are well suited to the reweighting strategy
employed in this analysis. They can occur when new resonances exhibit significant decay
branching ratios into $\WW$ and $\ZH$ while remaining consistent with the theoretical
bound: $\gstar < 4\pi$. Figure~\ref{fig:2Dfit-with-resonances} shows the allowed regions in
the $(\Wpar,\Ypar)$ plane for resonance mass hypotheses of 6 and 10\TeV. Due to the large 
widths considered, the low-mass tail has the potential to produce an observable effect 
within the visible mass range.

\section{Summary \label{sec:summary}}

A search for new physics effects due to Higgs boson compositeness in high-mass dielectron
and dimuon final states has been performed with the CMS detector, using LHC data collected
at center-of-mass energies of 13 and 13.6\TeV corresponding to a total integrated
luminosity of $207\fbinv$. The presence of such effects could imply the existence of new
massive boson resonances with couplings to difermions and dibosons. Assuming new boson
states with masses well above the \TeVns scale, stringent bounds are set on the oblique
parameters $\Wpar$ and $\Ypar$, which are zero in the standard model. These results are
also combined with previous CMS constraints on $\Wpar$ from $\Wprime$ searches to derive
the most stringent limits to date. One-dimensional fits, in which only one of the oblique
parameters is permitted to deviate from zero, yield the following limits: $-3.9\times
10^{-4} < \Ypar <3.5\times 10^{-4}$, and $-2.1\times 10^{-4} < \Wpar <1.6\times 10^{-4}$ at
95\% confidence level. Universal coupling scenarios with $\Wpar, \Ypar > 0$, such as
Strongly Interacting Light Higgs models, are excluded for $\Wpar > 7.8 \times 10^{-5}$ or
$\Ypar > 7.6 \times 10^{-4}$ at a credibility level above $95\%$. Taking into account the
current CMS bounds on the Higgs boson signal strength modifier~\cite{CMS:HIGGSNEW}, Higgs
boson compositeness scale limits at the \TeVns scale are set in Y-universal and Little Higgs
scenarios, independently of the value of the compositeness coupling strength $\gstar$.
Limits are also set for $\Zprime$ resonances in scenarios with large decay widths that
would result from significant decay branching fractions into diboson final states. In this
context, cases with large resonance decay widths are explored for the first time.

\begin{acknowledgments}
We congratulate our colleagues in the CERN accelerator departments for the excellent performance of the LHC and thank the technical and administrative staffs at CERN and at other CMS institutes for their contributions to the success of the CMS effort. In addition, we gratefully acknowledge the computing centers and personnel of the Worldwide LHC Computing Grid and other centers for delivering so effectively the computing infrastructure essential to our analyses. Finally, we acknowledge the enduring support for the construction and operation of the LHC, the CMS detector, and the supporting computing infrastructure provided by the following funding agencies: SC (Armenia), BMFWF and FWF (Austria); FNRS and FWO (Belgium); CNPq, CAPES, FAPERJ, FAPERGS, and FAPESP (Brazil); MES and BNSF (Bulgaria); CERN; CAS, MoST, and NSFC (China); MINCIENCIAS (Colombia); MSES and CSF (Croatia); RIF (Cyprus); SENESCYT (Ecuador); ERC PRG and PSG, TARISTU24-TK10 and MoER TK202 (Estonia); Academy of Finland, MEC, and HIP (Finland); CEA and CNRS/IN2P3 (France); SRNSF (Georgia); BMFTR, DFG, and HGF (Germany); GSRI (Greece); MATE and NKFIH (Hungary); DAE and DST (India); IPM (Iran); SFI (Ireland); INFN (Italy); MSIT and NRF (Republic of Korea); MES (Latvia); LMTLT (Lithuania); MOE and UM (Malaysia); BUAP, CINVESTAV, CONACYT, LNS, SEP, and UASLP-FAI (Mexico); MOS (Montenegro); MBIE (New Zealand); PAEC (Pakistan); MSHE, NSC, and NAWA (Poland); FCT (Portugal); MESTD (Serbia); MICIU/AEI and PCTI (Spain); MOSTR (Sri Lanka); Swiss Funding Agencies (Switzerland); NSTC (Taiwan); MHESI (Thailand); TUBITAK and TENMAK (T\"{u}rkiye); NASU (Ukraine); STFC (United Kingdom); DOE and NSF (USA).

\begin{sloppypar}
\setlength\emergencystretch{\hsize}
\hyphenation{Rachada-pisek} Individuals have received support from the Marie-Curie program and the European Research Council and Horizon 2020 Grant, contract Nos.\ 675440, 724704, 752730, 758316, 765710, 824093, 101115353, 101002207, 101001205, and COST Action CA16108 (European Union); the Leventis Foundation; the Alfred P.\ Sloan Foundation; the Alexander von Humboldt Foundation; the Science Committee, project no. 22rl-037 (Armenia); the Fonds pour la Formation \`a la Recherche dans l'Industrie et dans l'Agriculture (FRIA) and Fonds voor Wetenschappelijk Onderzoek contract No. 1228724N (Belgium); the Beijing Municipal Science \& Technology Commission, No. Z191100007219010, the Fundamental Research Funds for the Central Universities, the Ministry of Science and Technology of China under Grant No. 2023YFA1605804, the Natural Science Foundation of China under Grant No. 12535004, and USTC Research Funds of the Double First-Class Initiative No.\ YD2030002017 (China); the Ministry of Education, Youth and Sports (MEYS) of the Czech Republic; the Shota Rustaveli National Science Foundation (Georgia); the Deutsche Forschungsgemeinschaft (DFG), among others, under Germany's Excellence Strategy -- EXC 2121 ``Quantum Universe" -- 390833306, and under project number 400140256 - GRK2497; the Hellenic Foundation for Research and Innovation (HFRI), Project Number 2288 (Greece); the Hungarian Academy of Sciences, the New National Excellence Program - \'UNKP, the NKFIH research grants K 131991, K 138136, K 143460, K 143477, K 147557, K 146913, K 146914, K 147048, TKP2021-NKTA-64, 2026-4.1.1-MISSZI\'{O}K-2026-00005 and 2025-1.1.5-NEMZ\_KI-2025-00004, and MATE KKP and KKPCs Research Excellence and Flagship Research Groups grants (Hungary); the Council of Science and Industrial Research, India; ICSC -- National Research Center for High Performance Computing, Big Data and Quantum Computing, FAIR -- Future Artificial Intelligence Research, and CUP I53D23001070006 (Mission 4 Component 1), funded by the NextGenerationEU program, the Italian Ministry of University and Research (MUR) under Bando PRIN 2022 -- CUP I53C24002390006, PRIN PRIMULA 2022RBYK7T (Italy); the Latvian Council of Science; the Ministry of Science and Higher Education, project no. 2022/WK/14, and the National Science Center, contracts Opus 2021/41/B/ST2/01369, 2021/43/B/ST2/01552, 2023/49/B/ST2/03273, and the NAWA contract BPN/PPO/2021/1/00011 (Poland); the Funda\c{c}\~ao para a Ci\^encia e a Tecnologia (Portugal); the National Priorities Research Program by Qatar National Research Fund;  MICIU/AEI/10.13039/501100011033, ERDF/EU, ``European Union NextGenerationEU/PRTR", projects PID2022-142604OB-C21, PID2022-139519OB-C21, PID2023-147706NB-I00, PID2023-148896NB-I00, PID2023-146983NB-I00, PID2023-147115NB-I00, PID2023-148418NB-C41, PID2023-148418NB-C42, PID2023-148418NB-C43, PID2023-148418NB-C44, PID2024-158190NB-C22, RYC2021-033305-I, RYC2024-048719-I, CNS2023-144781, CNS2024-154769 and Plan de Ciencia, Tecnolog{\'i}a e Innovaci{\'o}n de Asturias, Spain; the Chulalongkorn Academic into Its 2nd Century Project Advancement Project, the National Science, Research and Innovation Fund program IND\_FF\_68\_369\_2300\_097, and the Program Management Unit for Human Resources \& Institutional Development, Research and Innovation, grant B39G680009 (Thailand); the Eric \& Wendy Schmidt Fund for Strategic Innovation through the CERN Next Generation Triggers project under grant agreement number SIF-2023-004; the Kavli Foundation; the Nvidia Corporation; the SuperMicro Corporation; the Welch Foundation, contract C-1845; and the Weston Havens Foundation (USA).
\end{sloppypar}
\end{acknowledgments}\section*{Data availability} Release and preservation of data used by the CMS Collaboration as the basis for publications is guided by the  \href{https://doi.org/10.7483/OPENDATA.CMS.1BNU.8V1W}{CMS data preservation, re-use and open access policy}.

\bibliography{auto_generated} 
\cleardoublepage \appendix\section{The CMS Collaboration \label{app:collab}}\begin{sloppypar}\hyphenpenalty=5000\widowpenalty=500\clubpenalty=5000\cmsinstitute{Yerevan Physics Institute, Yerevan, Armenia}
{\tolerance=6000
A.~Gevorgyan\cmsorcid{0000-0003-2751-9489}, A.~Hayrapetyan, V.~Makarenko\cmsorcid{0000-0002-8406-8605}, A.~Tumasyan\cmsAuthorMark{1}\cmsorcid{0009-0000-0684-6742}
\par}
\cmsinstitute{Institut f\"{u}r Hochenergiephysik, Vienna, Austria}
{\tolerance=6000
P.S.~Hussain\cmsorcid{0000-0002-4825-5278}, M.~Sonawane\cmsorcid{0000-0003-0510-7010}
\par}
\cmsinstitute{Marietta Blau Institute for Particle Physics, Vienna, Austria}
{\tolerance=6000
W.~Adam\cmsorcid{0000-0001-9099-4341}, L.~Benato\cmsorcid{0000-0001-5135-7489}, T.~Bergauer\cmsorcid{0000-0002-5786-0293}, M.~Dragicevic\cmsorcid{0000-0003-1967-6783}, S.~Gundacker\cmsorcid{0000-0003-2087-3266}, A.K.~Guven\cmsorcid{0009-0004-5670-5138}, M.~Jeitler\cmsAuthorMark{2}\cmsorcid{0000-0002-5141-9560}, N.~Krammer\cmsorcid{0000-0002-0548-0985}, A.~Li\cmsorcid{0000-0002-4547-116X}, D.~Liko\cmsorcid{0000-0002-3380-473X}, M.~Matthewman, A.~Pfeiffer\cmsorcid{0000-0001-5328-448X}, J.~Schieck\cmsAuthorMark{2}\cmsorcid{0000-0002-1058-8093}, R.~Sch\"{o}fbeck\cmsAuthorMark{2}\cmsorcid{0000-0002-2332-8784}, M.~Shooshtari\cmsorcid{0009-0004-8882-4887}, N.~Van~Den~Bossche\cmsorcid{0000-0003-2973-4991}, W.~Waltenberger\cmsorcid{0000-0002-6215-7228}, C.E.~Wulz\cmsAuthorMark{2}\cmsorcid{0000-0001-9226-5812}
% ...
\par}
$^{1}$Also at Yerevan State University, Yerevan, Armenia\\
$^{2}$Also at Technische Universit\"{a}t Wien, Vienna, Austria\\
% ...
\end{sloppypar}
\end{document}